\documentclass[longauthor, twocolumn, twocolappendix]{aastex631} 

\newcommand{\qmin}{\mathcal{Q}_{y}}

\newcommand{\mWD}{m_{\scriptsize\textsc{wd}}}
\newcommand{\mPR}{m_{\scriptsize\textsc{pr}}}
\newcommand{\mRG}{m_{\scriptsize\textsc{rg}}}
\newcommand{\qL}{q_{\scriptsize\textsc{l}}}
\newcommand{\qH}{q_{\scriptsize\textsc{h}}}
\newcommand{\qB}{q_{\scriptsize\textsc{b}}}
\newcommand{\msun}{\small \rm M_\odot}

\newcommand{\Pwd}{{\rm P}_{\scriptsize\textsc{wd}}}

\usepackage{amsmath}

\begin{document}

\title{The White Dwarf Pareto: Tracing Mass Loss in Binary Systems}

\correspondingauthor{Sahar Shahaf}
\email{sahar.shahaf@weizmann.ac.il}

\author[0000-0001-9298-8068]{Sahar Shahaf}
\affiliation{Department of Particle Physics and Astrophysics\\ Weizmann Institute of Science\\ Rehovot 7610001, Israel}

\begin{abstract}
The white dwarf mass distribution has been studied primarily at two extremes: objects that presumably evolved as single stars and members of close binaries that likely underwent substantial interaction. This work considers the intermediate separation regime of ${\sim}\,1$ au and demonstrates how binary interaction affects white dwarf masses. The binary mass ratio distribution is utilized for this purpose. Modeled as a truncated Pareto profile, this distribution provides insights into the populations' properties and evolutionary history. When applied to homogeneous samples of binaries with giant primaries of similar age, the distribution’s shape constrains the fraction of white dwarf companions, the white dwarf mass distribution, and the properties of their progenitors. As a test case, this method is applied to a small spectroscopic sample of binaries in open clusters with red giant primaries and orbital periods between $0.5$ and $20$ years. The analysis reveals that white dwarfs in these systems are ${\sim}\,20\%$ less massive than their isolated counterparts, with a typical mass of ${\sim}\,0.55\,\msun$. Their progenitors likely lost $80-85\%$ of their mass, with binary interactions enhancing mass loss by an additional  ${\sim}\,0.2\,\msun$. These findings highlight the utility of this approach for studying binary evolution and improving population models, particularly with future datasets from Gaia and other large-scale surveys.
\end{abstract}

\section{Introduction} \label{sec:intro}

Stellar multiplicity is established during the early stages of star formation, with primordial configurations dynamically evolving to produce the observed populations of single, binary, and higher-order multiple systems \citep{larson72,reipurth14, beuther19, rosen20, Offner2023}. Binary systems are particularly prevalent, comprising about half of all Sun-like stars, three-quarters of A-type stars, and nearly all B- and O-type stars \citep[e.g.,][]{raghavan10, Duchene2013, Offner2023}. The high occurrence of binaries significantly influences various astrophysical processes, a fact recognized even before large samples became available, motivating early empirical studies of binary populations over a century ago \citep[e.g.,][]{biesbroeck1916,opik24, kuiper35}.

Whether single or part of a binary system, all stars eventually exhaust their nuclear energy supplies, leaving behind stellar remnants, most commonly white dwarfs \citep{Althaus2010}. Despite the high frequency of white dwarfs, with hundreds of thousands already discovered either in binaries or isolation \citep{GentileFusillo2021, Shahaf2024}, key aspects of their population remain incompletely understood \citep[e.g.,][]{Isern2022, Hallakoun2024}. In particular, the relationship between a white dwarf’s mass and that of its progenitor star, known as the initial-to-final mass relation, or the IFMR, remains an area of active research \citep{tremblay16, cummings16, cummings18, El-Badry2018, Cunningham2024, Hollands2024,  Addari2024}. 

The extent to which binarity affects the mass of white dwarfs strongly depends on the orbital separation. At sufficiently large separations, the two stars are expected to evolve independently \citep{Barrientos2021, Hollands2024, Heintz2022}. However, at close separation, binary interaction plays an important role. It becomes evident when considering the population of extremely low-mass  \citep[e.g.,][]{Brown2010, Kosakowski2023}, ultramassive \citep{Cheng2020, Fleury2022, Kilic2023} or highly magnetized white dwarfs \citep{Moss2023}. However, a significant portion of the population presumably exists in neither of these extremities. Still, until recently, white dwarfs in binaries separated by ${\sim}1$~au have eluded detection \citep{Holberg2013}. 

This situation has changed with the recent data release by the Gaia spacecraft \citep{El-Badry2024, Tremblay2024}. The astrometric and spectroscopic orbits from Gaia uncovered a large population of binaries comprised of a white dwarf and main sequence star at this intermediate separation regime \citep{Shahaf2024, Yamaguchi2024b, Yamaguchi2024}. Some of the emerging properties of this population already provided new insights and challenges to stellar and binary evolution models \citep{Hallakoun2024, Shahaf2024, Yamaguchi2024b, Rekhi2024}. 

This work presents a method to estimate the white dwarf mass distribution and constrain the IFMR using binaries where the primary star is a red giant. A truncated power-law (i.e., a Pareto distribution) approximates the mass ratio distribution, whose properties are sensitive to the IFMR parameters. This approach can be applied to spectroscopic binary samples, allowing exploration across a wide range of orbital separations and offering new insights into binary and stellar evolution in regimes that have been previously underexplored.

The structure of the paper is as follows: Section~\ref{sec: companions of giants} presents the expected shape of the mass ratio and modified mass function distributions; Section~\ref{sec: merm07 sample} applies this model, as a test case, to a spectroscopic sample of binaries with red giant primaries from open clusters. The results are discussed in Section~\ref{sec: discussion}. Section~\ref{sec: summary} summarizes the findings and outlines avenues for future study.

\section{Companions of giant stars}
\label{sec: companions of giants}
\subsection{The white dwarf fraction}
\label{sec: wd fraction}
 
Consider an open star cluster of age $\tau$. The cluster's turnoff mass, $m_\tau$, defines the threshold above which stars have evolved off the main sequence. Now, suppose a binary system is identified in this cluster, where the primary star is a red giant of mass $\mRG$, and the secondary is too faint to be directly identified. The mass of the secondary can be constrained indirectly from the orbital motion of the primary. However, determining its exact nature based solely on the orbital elements and primary mass is challenging. 
For example, without additional information, it is difficult to determine whether a ${\sim}\,0.6~\msun$ companion of a red giant is a white dwarf or a late-type main sequence star \citep[e.g.,][]{Shahaf2019b}. Nevertheless, compared to unequivocal classification, statistical estimates are often more feasible. 

The occurrence rate of white dwarf companions stems from the properties of the primordial, zero-age main sequence population. To model it, assume that the distribution of primordial primary masses, $m_1$, is similar to the \citet{salpeter55} initial mass function,  where $\alpha$ signifies the exponential index of the distribution. The primordial mass ratio distribution is taken to be flat so that the secondary mass, $m_2$, is uniformly distributed between $0$ and $m_1$. These assumptions yield a bivariate density function,
\begin{equation}
\Phi(m_1,m_2) \propto m_1^{-(\alpha+1)}
\label{eq:Pm1m2}
\end{equation}
if $0 < m_2 \leq m_1$ and zero otherwise. The impact of deviations from uniformity on the inferred properties of the sample is discussed below.

To assess the probability of the secondary companion being a white dwarf, we consider two possible scenarios: first, that the present-day giant star is also the primordial primary, and second, that it is the primordial secondary. In the first scenario, $m_1$ can be identified with $\mRG$, assuming insignificant mass loss. Equation~(\ref{eq:Pm1m2}) suggests the fraction of systems in this state, ${\rm P}_{\scriptsize\textsc{ms}}$, is proportional to $\mRG^{-\alpha}$. In this case, the secondary companion is most likely a main-sequence star. Other, less common, evolutionary scenarios are also possible but presumably occur much less frequently.

In the second scenario, the faint star was originally more massive and has already evolved into a compact object. Most likely a white dwarf. This requires the primordial primary to have been massive enough to complete its post-main-sequence evolution, i.e., $m_1 > m_\tau + \Delta$, where $\Delta$ is the gap between the turn-off mass and the least massive white dwarf progenitor star. Since the red giant was originally the primordial secondary, $\mRG \simeq m_2$. This primordial secondary is now a red giant, therefore, we can place bounds on its mass:  $m_\tau < m_2 < m_\tau + \Delta$. From equation~(\ref{eq:Pm1m2}), the fraction of systems with white dwarf companions, $\Pwd$, is approximately proportional to $\alpha^{-1}(m_\tau + \Delta)^{-\alpha}$.

The odds ratio between the two cases does not depend on the proportion coefficients, 
\begin{equation}
    {\rm O}_{\scriptsize\textsc{wd}} \equiv \frac{{\rm P}_{\scriptsize\textsc{wd}}}{{\rm P}_{\scriptsize\textsc{ms}}}\simeq \frac{1}{\alpha} 
    \bigg(\frac{\mRG}{m_\tau}\bigg)^{\alpha}
    \bigg( 1+  \frac{\Delta}{m_\tau} \bigg)^{-\alpha}.
\label{eq:Owd}
\end{equation}
In the following, it is assumed that the contribution of other configurations, such as stripped stars, triple systems, or neutron stars, is small. Hence, $\Pwd+ {\rm P}_{\scriptsize\textsc{ms}} \simeq 1$ and the fraction of white dwarf companions is
\begin{equation}
    \Pwd \simeq \frac{{\rm O}_{\scriptsize\textsc{wd}}}{1+{\rm O}_{\scriptsize\textsc{wd}}}\,.
    \label{eq:Pwd}
\end{equation}
Next, we use this expression to describe the mass ratio distribution of the sample.

\subsection{The mass ratio distribution}
The mass ratio of a binary system, $q$, is defined as the mass of the faint secondary component, whether a main-sequence star ($m_{\scriptsize\textsc{ms}}$) or a white dwarf ($\mWD$), divided by the mass of the luminous red giant primary, $\mRG$.

The sample’s mass ratio distribution consists of two components: one for the sub-population in which the red giant is the primordial primary and another where it is the primordial secondary. In the first case, assuming negligible mass loss or transfer during the giant’s evolution, the mass ratio distribution should remain flat. Alternatively, if the red giant was the primordial secondary, the white dwarf progenitor lost a substantial fraction of its mass, changing the distribution's shape. Therefore, the resulting mass ratio distribution can be expressed as
\begin{equation} 
    f_{q} = (1- \Pwd) + \Pwd \cdot f_{q}^{\scriptsize\textsc{wd}} \,.
    \label{eq: MRDwdms} 
\end{equation} 
The first term represents the uniformly distributed systems from the first scenario, and the second term represents systems where the companion has evolved into a white dwarf. To fully describe the mass ratio distribution, it is necessary to define $f_{q}^{\scriptsize\textsc{wd}}$. 

The white dwarf progenitor is initially the primordial primary star, with a mass denoted as $\mPR$ (equivalent to $m_1$). Consequently, the white dwarf mass distribution follows a truncated power law with the same exponential index as the mass distribution from equation~(\ref{eq:Pm1m2}), namely, it is proportional to $\mPR^{-(\alpha +1)}$. This power-law distribution is truncated by the Chandrasekhar limit from above and by the remnant mass of the least massive progenitor (approximately $m_\tau + \Delta$; see above) from below. The progenitor’s mass relates to that of the white dwarf through the IFMR, which, for simplicity, is modeled as a linear relation,
\begin{equation}
    \mWD = A\,\mPR + B.
    \label{EQ: IFMR main}
\end{equation}
Using the inverse relation, it follows that the $\mWD$ distribution is proportional to $(\mWD - B)^{-(\alpha +1)}$.

Consider the mass ratio distribution of a binary system with a red giant of some fixed mass, $\mRG$. In this case, the mass ratio distribution inherits the truncated power-law form. Probability density functions of this type are often referred to as \textit{generalized Pareto distributions}. The distribution is characterized by the initial mass function exponent, $\alpha$, and three additional parameters: $\qB$, $\qL$, and $\qH$. The resulting distribution is 
\begin{equation}
    f_{q}^{\scriptsize\textsc{wd}} = 
        \zeta_0 \, \big(q -  \qB\big)^{-(\alpha+1)}
    \label{eq:fWD}  
\end{equation}
if $q\in (\qL, \qH)$ and zero otherwise. The mass-ratio parameters satisfy $\qB<\qL<\qH$, and $\zeta_0$ is a normalization constant, where
\begin{equation*}
    \zeta_0  = \alpha~\frac{(\qH-\qB)^\alpha (\qL-\qB)^\alpha}{(\qH-\qB)^\alpha - (\qL-\qB)^\alpha } \,.
\end{equation*}
Notably, the mass of the giant and evolutionary phase do not appear explicitly in equation~(\ref{eq:fWD}). Therefore, the requirement that the primary star is a red giant can be relaxed. This will be discussed later when prospective samples are considered for analysis.
%
\begin{figure}
	\centering
	\includegraphics[width=0.975\linewidth ,trim={0 0 0 0 },clip]{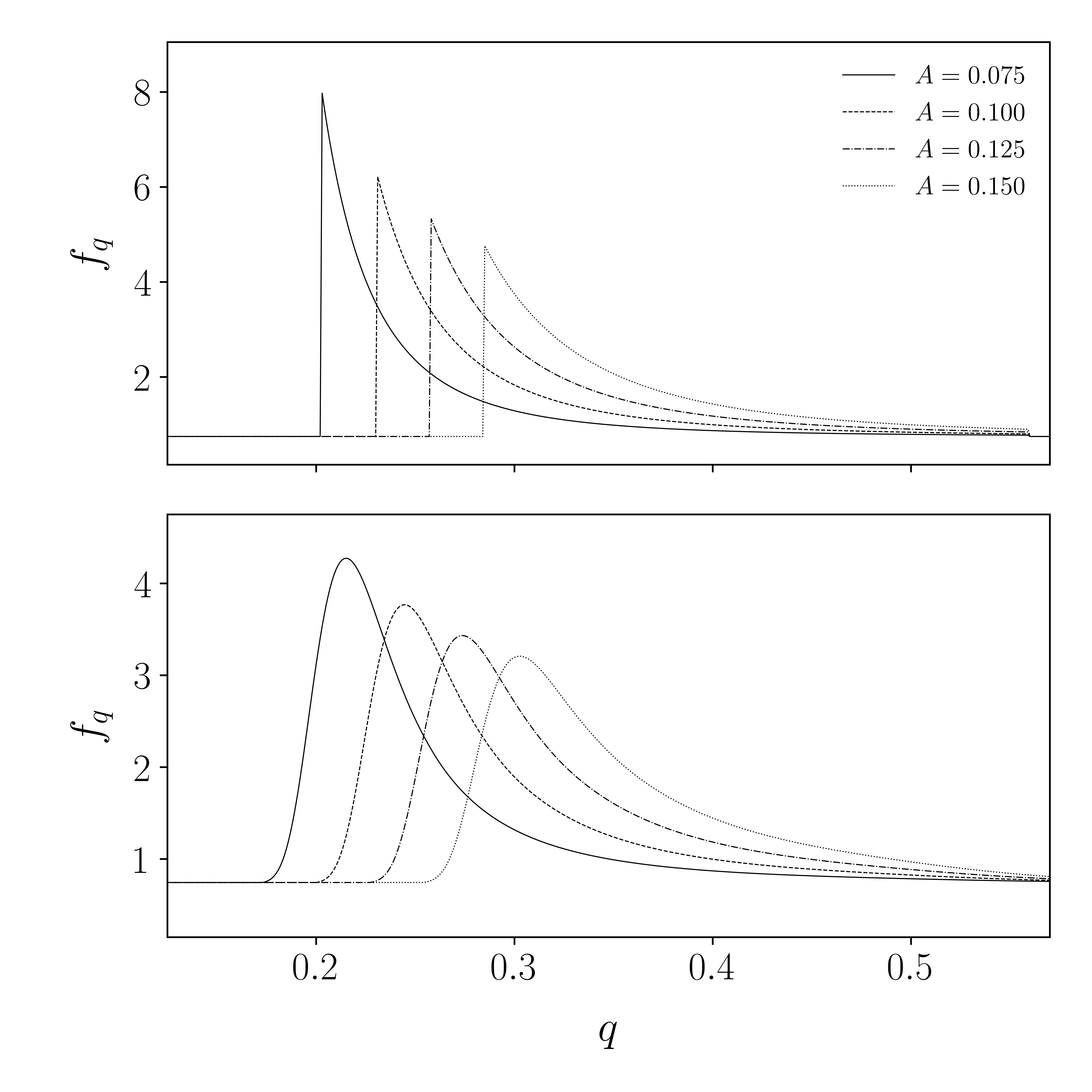}
	\caption{ \textit{Top panel} -- The mass ratio distribution from equation~(\ref{eq: MRDwdms}), for several IFMR slopes (see legend) and a fixed intercept ($B$) of $0.3$~${\rm M}_\odot$. The distributions were calculated assuming a red giant mass of $2.50~{\rm M}_\odot$, turnoff mass of $2.25$~${\rm M}_\odot$, and $\Delta=0.5$~${\rm M}_\odot$. The initial mass function exponent used is $\alpha=2.35$.  \textit{Bottom panel} -- the corresponding marginal mass ratio distribution derived assuming the uncertainty in the giant's mass is $0.25$~${\rm M}_\odot$, and $m_\tau=\mRG-\Delta/2$. Line styles represent the same parameters as in the top panel.}
	\label{fig:MRDmodelExample}
\end{figure}
%

The distribution parameters are related to the IFMR from equation~(\ref{EQ: IFMR main}) through the following relations:
\begin{equation}
    \begin{aligned}
        \qB &\simeq B/\mRG, \\
        \qL &\simeq A\big({m_\tau} + {\Delta}\big)/{\mRG}+ \qB, \quad \text{and}\\
        \qH &\simeq {1.4 \, \msun}/{\mRG}.
    \end{aligned}
    \label{eq:qBounds}
\end{equation}
The top panel of Figure~\ref{fig:MRDmodelExample} illustrates the shape of the obtained mass ratio distribution. In practice, the giant's mass, $\mRG$, is estimated up to some uncertainty,  $\delta m$. This uncertainty should soften the sharp features of the distribution, as it affects the cut-off values from equation~(\ref{eq:qBounds}). Therefore, averaging out the giant's mass could provide a more realistic distribution visualization. This approach is demonstrated in the bottom panel of Figure~\ref{fig:MRDmodelExample}, where the giant's mass is assumed to be normally distributed.

The peak of the distribution originates from the lower cut-off mass ratio, $q_{\scriptsize\textsc{l}}$. While marginalizing over the giant's mass smooths the distribution, the peak remains a prominent and localized feature. Since $q_{\scriptsize\textsc{l}}$ is sensitive to the parameters of the IFMR, so is the position of the peak, $q_{\rm p}$. Therefore, it could provide immediate insights upon inspection. As a crude approximation, we take 
\begin{equation}
\label{eq: qp_to_IFMR}
   q_{\rm p} \approx A\big(1 + \varepsilon \big) + q_{\scriptsize\textsc{b}}\,, 
\end{equation}
where $\varepsilon$ is on the order of $10^{-1}$, and the approximation accuracy degrades roughly like $(\delta m\big/m)^2$. 
Figure~\ref{fig:fqpeak} illustrates the peak's position as a function of the slope and intercept of the IFMR.

\subsection{The modified mass function distribution}
\label{sec: mmf}
Empirical studies of spectroscopic binaries can, in principle, constrain the white dwarf IFMR across a wide range of orbital separations and evolutionary scenarios. However, the red giant dominates the system's emitted flux in most cases. Thus, only the spectral lines originating from the giant are detected, and only the giant's velocities are measured. As a result, the mass ratio, $q$, and orbital inclination, $i$, cannot be independently derived.

However, if sufficient radial velocity measurements are obtained, the orbital period, radial velocity semi-amplitude, and eccentricity can be fitted to reproduce the observed orbital modulation. The mass function, $f_m$, can be derived from these orbital parameters. The ratio between $f_m$ and $\mRG$ often called the reduced mass function, can be expressed as a combination of the mass ratio and orbital inclination:
\begin{equation} \label{EQ: reduced mass function}
y \equiv \frac{f_m}{\mRG}  =  \frac{q^3}{(1+q)^2} \,  \sin^3i \,.
\end{equation}
Notably, each value of $y$ is associated with a minimal possible mass ratio, $\qmin$, which is determined by setting the inclination to $90^{\circ}$.

The reduced mass function distribution is obtained by assuming that the binary's orbital plane is randomly oriented. However, deriving the underlying mass ratio distribution from an observed set of reduced mass function values is not straightforward \citep[e.g.,][]{mg92, boffin93, heacox95}. To address this challenge, \citet{shahaf17} introduced the {\it modified mass function},  
\begin{equation} \label{EQ:S}
S \equiv 1 - \int_{\qmin}^{1}{\sqrt{1 - y^{2/3} \,{(1+q)^{4/3}}{q^{-2}}}\,dq} \,.
\end{equation}
%
Unlike the reduced mass function, the modified mass function distribution, $f_S$, qualitatively follows the shape of the underlying mass ratio distribution. The expression to derive $f_S$ from  $f_q$ is available in Appendix~\ref{app: mmf dist}.

\begin{figure}
	\centering	\includegraphics[width=0.975
    \linewidth ,trim={0 0 0 0 },clip]{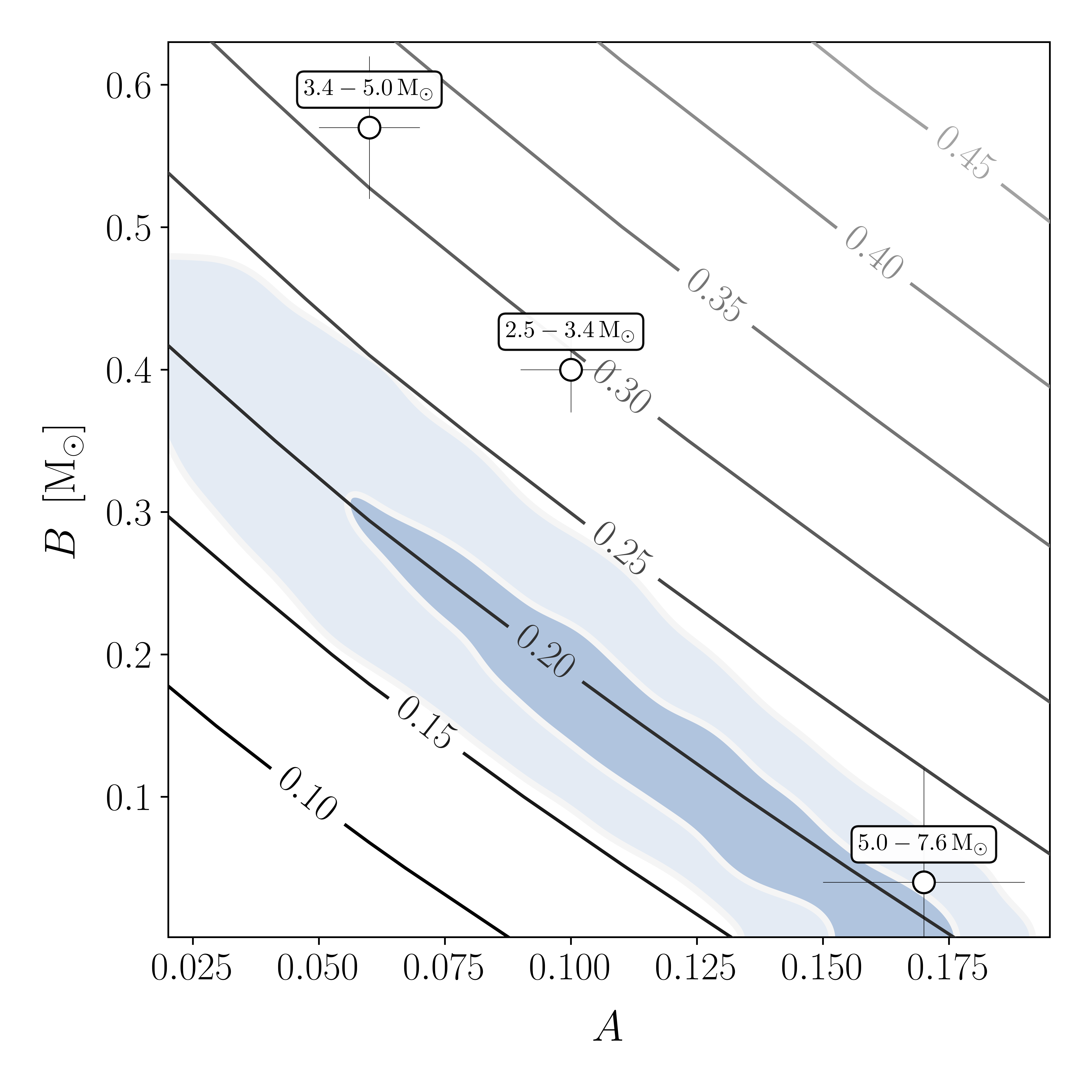}
	\caption{The position of the peak of the mass ratio distribution, as a function of the IFMR slope ($A$) and intercept ($B$).  The solid annotated lines depict the mass-ratio value at which the distribution attains its maximum. The distributions were calculated assuming $m_{\textsc{rg}}=2.4\pm0.4$ ${\rm M}_\odot$, $\Delta=0.5$ ${\rm M}_\odot$, $\alpha=2.35$. The open circles represent the IFMR parameters fitted by \citet[][]{Cunningham2024}. The light-blue shading represents the 1 and 2$\sigma$ levels of the derived IFMR (see Section~\ref{sec: merm07 sample}).}
	\label{fig:fqpeak}
\end{figure}

\begin{figure}
	\centering
	\includegraphics[width=0.975\linewidth ,trim={0 0 0 0 },clip]{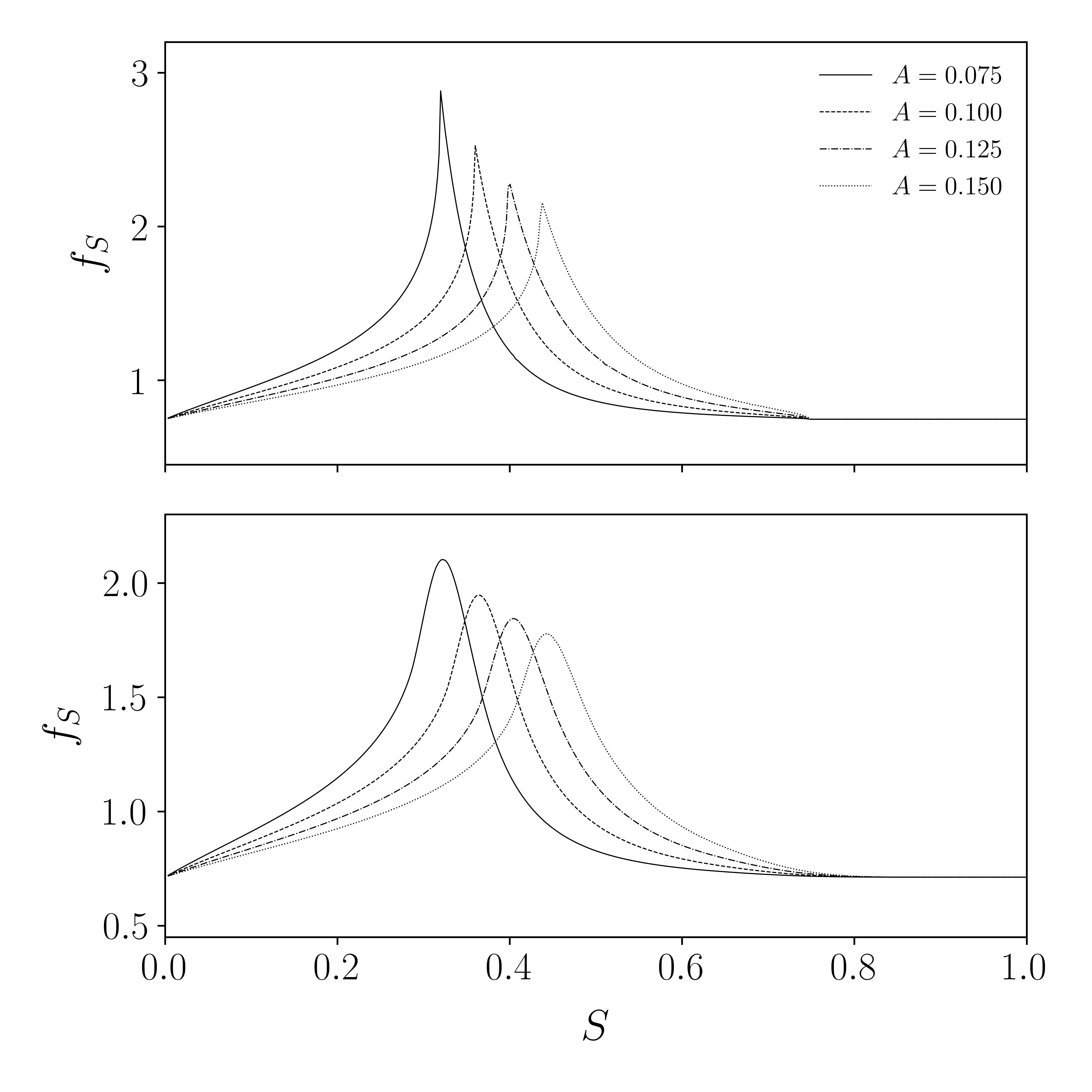}
	\caption{The modified mass function distributions for red giants in open clusters. The top and bottom panels plots were produced using the same parameters as in Figure~\ref{fig:MRDmodelExample}.}
	\label{fig:fS}
\end{figure}

The similarity between $f_S$ and $f_q$ can be used to constrain the IFMR. The top panel of Figure~\ref{fig:fS} presents the resulting distribution of the modified mass function for the same set of parameters used in Figure~\ref{fig:MRDmodelExample}. The relationship between the peak positions of the two distributions may be useful. The relation between the peak positions of $f_S$ and $f_q$ for the parameter space considered in Figure~\ref{fig:fqpeak} roughly follows the relation
\begin{equation}
\label{eq:lSlq}
    \log_{10}q_{\rm p} \simeq {0.20}\,\log_{10}^2{S_{\rm p}} + 1.30\log_{10}S_{\rm p}-0.08\,,
\end{equation}
where $q_{\rm p}$ and $S_{\rm p}$ are the values at which the mass ratio and modified mass function distributions attain their maximum.

\subsection{A Synthetic population}
\label{sec: popsyn}
We constructed a synthetic sample of binary stars to evaluate an approximated analytical expression for the mass ratio and the modified mass function distribution. This sample was generated using the \texttt{cogsworth}\footnote{Available online at \href{https://cogsworth.readthedocs.io}{cogsworth.readthedocs.io}.} package \citep{Wagg2024}, which integrates with the \texttt{COSMIC} population synthesis code \citep{Breivik2020}.

The synthetic population parameters were set to produce a simplistic scenario and aim to avoid common envelope evolution, significant mass transfer phases, or strong tidal interactions. The fiducial binary orbits were assumed circular, with periods uniformly distributed between 100 and 200 years. Primordial primary masses followed the \citet{salpeter55} initial mass function, while secondaries were drawn from a flat mass ratio distribution. A fixed metallicity of $0.2$~dex was applied across the population, and the white dwarf IFMR was modeled using the \citet{Han1995} relation.

The synthetic population evolved over 1~Gyr. To isolate binaries containing a red giant, we selected systems with a \texttt{COSMIC} \texttt{kstar} index of $2$, $3$, or $4$, corresponding to stars on the Hertzsprung gap, first giant branch, or helium-burning phase. We restricted the companions of these red giants to main-sequence stars (\texttt{kstar} $0$ or $1$), white dwarfs (\texttt{kstar} $10$, $11$, or $12$), or additional red giants of lower mass. To minimize the effects of binary interaction, evolved systems with orbital periods under 10 years were excluded. The simulation and these criteria yielded a synthetic sample of ${\sim}14,000$ binaries.

\begin{figure}
	\centering	\includegraphics[width=0.975\linewidth ,trim={0 0 0 0 },clip]{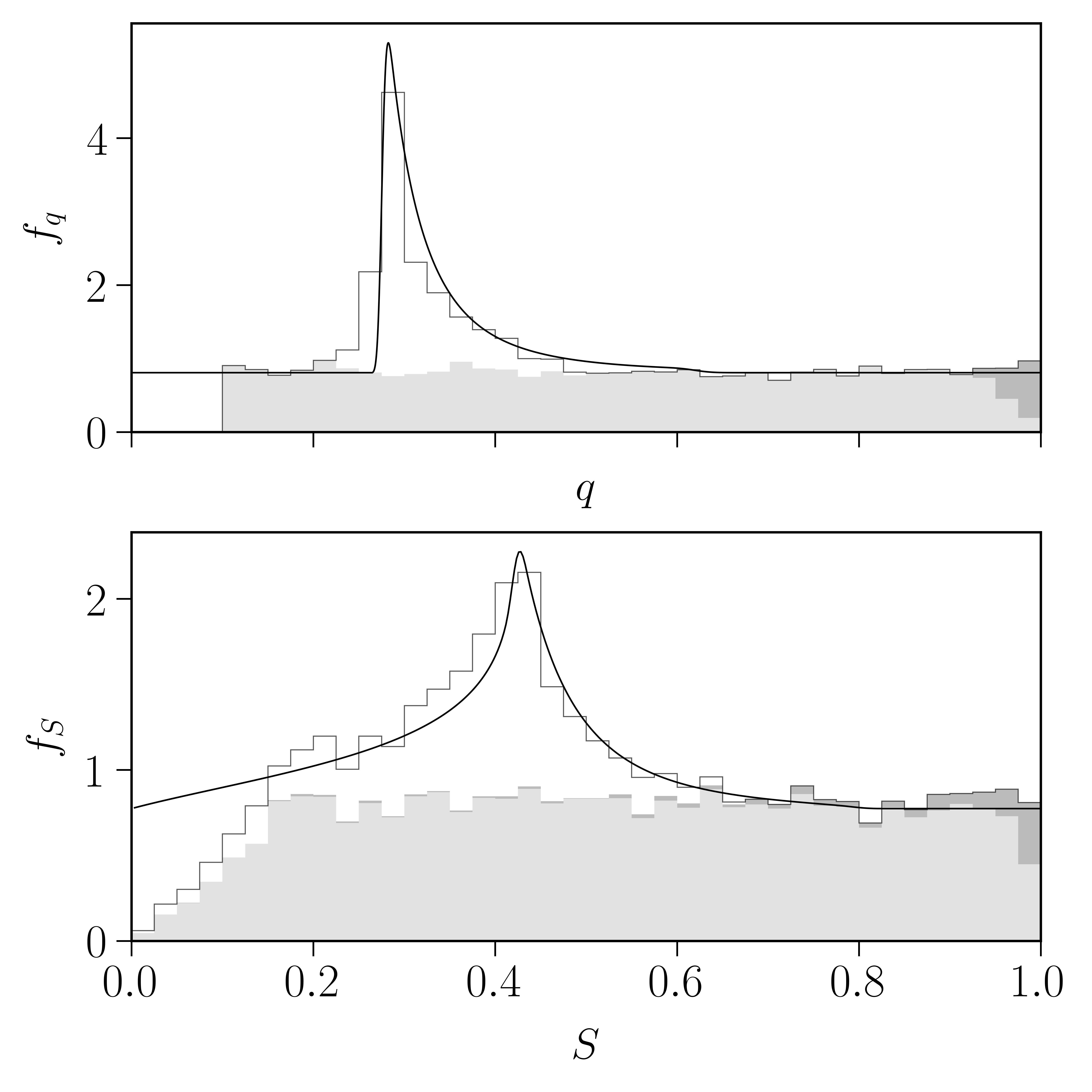}
	\caption{Mass ratio (top panel) and modified mass function (bottom panel) distribution of the synthetic population generated by \texttt{cogsworth}. The color of the histogram bin indicates the type of secondary companion: white, light gray, and dark gray bins represent white dwarfs, main sequence, and evolved giant secondaries, respectively. The thin black lines represent the expected shape of the distribution according to the analytical prescription (see text).}
	\label{fig:fq_synthetic}
\end{figure}

Figure~\ref{fig:fq_synthetic} presents histograms of the mass ratio and the modified mass function for the synthetic sample. The histograms distinguish subpopulations of giants paired with main-sequence stars, white dwarfs, or other giants. Most systems with main-sequence companions appear in light gray, while systems with white dwarf companions are shown as white bars. At near-unity mass ratios, the population predominantly comprises pairs of giant stars. The truncation at mass ratios below $0.1$ arises from the absence of sub-stellar companions in the simulation.

The expected $f_q$ and $f_S$ distributions are overlaid on the histograms in Figure~\ref{fig:fq_synthetic}. The red giant masses in the synthetic sample are not normally distributed. However, to align with prior discussions and figures, we assumed $\mRG$ is Gaussian and used the mean and standard deviation of the red giants in the synthetic sample ($2.25$ and $0.05$~$\msun$, respectively). The initial mass function exponent was set to the Salpeter value of $2.35$, and $\Delta$ was set to $0.25$ $\msun$.  We inspected $0.8{-}1$ Gyr Solar-metallicity PARSEC\footnote{PAdova and tRieste Stellar Evolution Code, version 1.2S, available online via \href{http://stev.oapd.inaf.it/cmd}{stev.oapd.inaf.it/cmd}.}  isochrones to justify the latter selection \citep[]{Bressan_2012, Chen_2014, Chen_2015, Tang_2014}. Finally, we adopted an IFMR of
\begin{equation*}
\mWD =  0.104 \, \mPR + 0.36 \,\, \msun.
\end{equation*}
This relation corresponds to one segment of the broken linear relation proposed by \citet{Han1995}, which best fits the distribution. The figure shows that the analytical prescription describes the synthetic distribution even without a detailed fitting procedure. This example shows how the equation~(\ref{eq: MRDwdms}) can describe the distribution if the IFMR is known. In Section~\ref{sec: merm07 sample}, we consider the inverse problem of inferring the IFMR from a set of observed binary orbits.

\subsection{Detection function}
The discussion above assumes an idealized scenario where all target population binaries are equally detectable, which is false in actual samples. However, a detection bias arising from finite instrumental precision can limit the detectability of systems with low mass ratios. Here, we consider a detection function, $\mathcal{D}$, for a sample of spectroscopic binaries, following the method of \citet{mg92}. 

The detection function is estimated by assuming that the survey is characterized by a radial velocity semi-amplitude threshold, $K_{\rm th}$, below which binaries remain undetected.  The orbital speed of a binary in a circular orbit is determined by its mass ratio, $q$, orbital period, $P$, and primary mass, $\mRG$. The detectability of a system is reduced to the span of orbital inclinations in which the radial component of the orbital speed exceeds the detection threshold. The critical inclination, $i_{\rm c}$, below which detection is impossible, is given by 
\begin{equation}
    \sin i_{\rm c} =  {K_{\rm th}} \bigg(\frac{1}{2\pi G}\frac{P}{ \mRG}\bigg)^{1/3} \frac{(1+q)^{2/3}}{q}\, .
\end{equation}
The detection function quantifies the detection probability. Assuming that the orbits are randomly oriented, 
\begin{equation}
\label{eq: detection function}
    \mathcal{D}(q) = \sqrt{1-\sin^2 i_{\rm c}}\,~.
\end{equation}
See \citet{mg92} for a detailed derivation of this relation. Examples can be found also in \citet{shahaf17} and \citet{Shahaf2019}.

The resulting correction function is a multiplicative factor imposed on the mass ratio distribution. The model for the observed biased distribution is, therefore, given by replacing the idealized model for $f_q$ from Equation~(\ref{eq: MRDwdms}) with the corrected version, namely
\begin{equation}
    f_q^{\rm obs} \propto f_q \times \mathcal{D}.
    \label{eq: fq_obs}
\end{equation}
This probability density function inherits the parameters of equations and ~(\ref{eq: MRDwdms}),~(\ref{eq:fWD}), and~(\ref{eq: detection function}). Since $f_q$ is normalized, the normalization constant depends on the parameters of the detection function. 

\section{A test case: The Mermilliod sample}
\label{sec: merm07 sample}
\subsection{Sample description}

In an extensive spectroscopic campaign spanning ${\sim}\,20$ years, \citet[]{mermilliod07} monitored $1309$ red giants across $187$ open clusters, achieving a typical radial velocity accuracy of ${\sim}\,0.4$ ${\rm km\,s}^{-1}$. They identified $289$ spectroscopic binaries and provided orbital solutions for $156$ of these systems. \citet{north14} estimated the mass ratio distribution of the binaries in this sample and identified a peak corresponding to secondaries of ${\sim}\,0.6$~$\msun$ and suggested it stems from a population of white dwarf companions. North also noted that their mass distribution seemed inconsistent with the expected IFMR.

\begin{figure}
	\centering
	\includegraphics[width=0.975\linewidth ,trim={0 0 0 0 },clip]{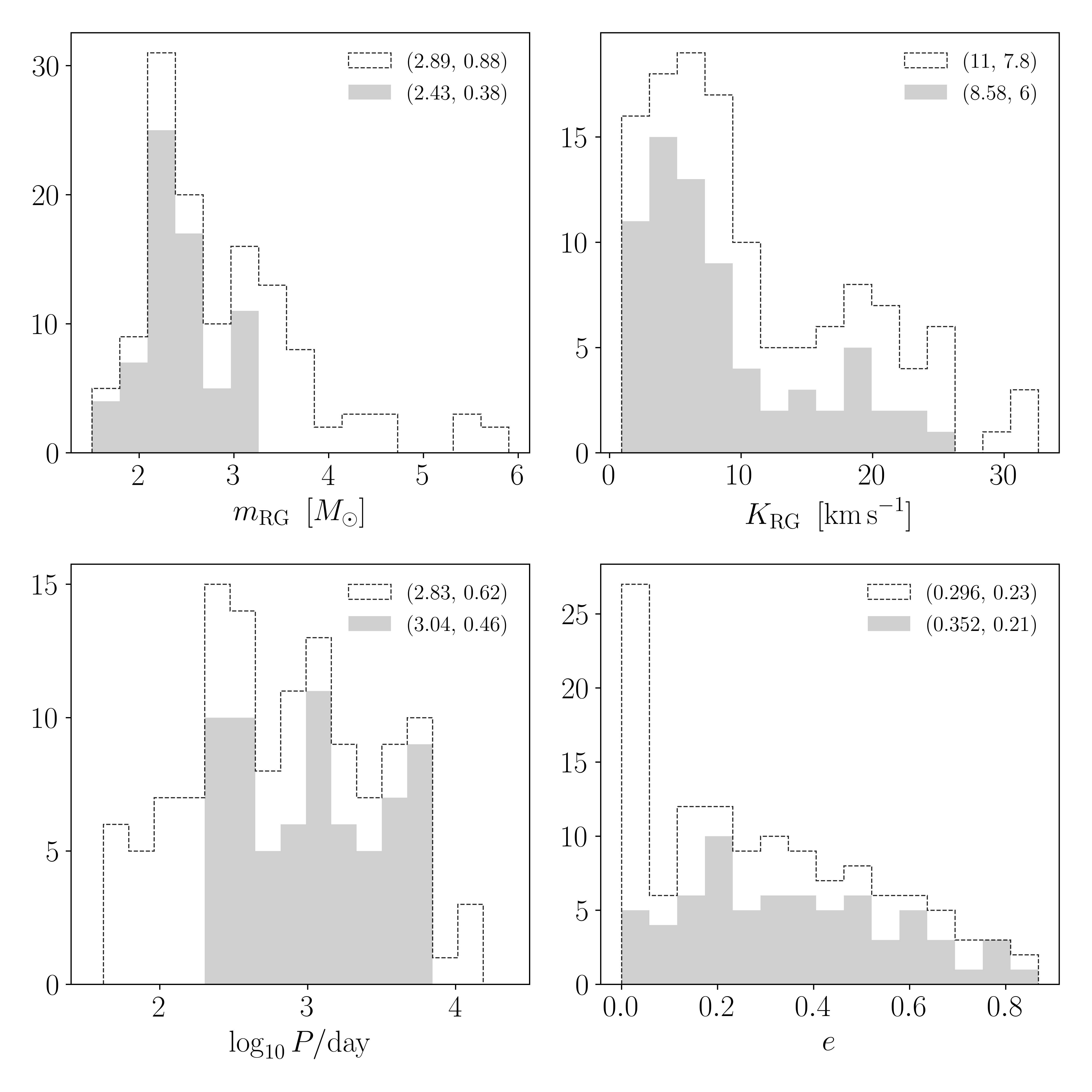}
	\caption{The properties of the sample. Clockwise from the top-left panel: histograms of the red giant mass, radial velocity semi amplitude, orbital eccentricity, and logarithm of the orbital period. The grey bars show the distribution selected subsample of $69$ systems and the entire sample of $125$ binaries is overlaid as a black dashed line. The legends of each panel provide the mean and standard deviation of each quantity. 
 }
	\label{fig:orbhist}
\end{figure}

In a subsequent study, \citet{swaelmen17} validated the cluster membership of the binaries in this sample. After filtering out systems with low cluster membership probability, they provided reliable mass estimates for $125$ red giants and derived the mass ratio distribution of the sample. Their finding corroborated the results of \citet{north14}. The full list, including mass estimates, is available online (see table A.4 therein). We select a subsample of $69$ systems with red giant primaries less massive than $3.25$~$\msun$ and orbital periods between $0.5$ and $20$ years. Figure~\ref{fig:orbhist} shows histograms of red giant mass, radial velocity semi-amplitude, logarithm of orbital period, and orbital eccentricity. 

The binaries in the selected sample are associated with $21$ different open clusters, listed in a supplementary table. The clusters have ages around $0.83^{+0.73}_{-0.24}$ Gyr and near-solar metallicities of $[{\rm Fe}/{\rm H}] \simeq 0.0^{+0.2}_{-0.1}$ dex \citep{Dias2021, Hunt2023}. 
Two systems have minimum mass ratios above one: star 4 in NGC 1528 and star 70 in NGC 6633, with $\qmin$ of $1.10\pm0.05$ and $1.07\pm0.02$, respectively. As demonstrated above, some binaries with mass ratios close to unity are expected to comprise two evolved stars. These systems are, therefore, retained and treated as equal mass binaries (modified mass function set to one).

Figure~\ref{fig:merm07_mmfhist} shows a histogram of the sample's modified mass function values. Using the rightmost histogram bins, we estimate $\Pwd\approx0.25$ and estimate that the peak of the distribution is positioned at $S_p\approx0.29$. Equation~(\ref{eq:lSlq}) identifies this peak with $q_p\approx0.185$, suggesting that $A\sim0.2$, assuming the intercept value is negligible. With these preliminary estimates, we now infer the distribution's parameters. 

\subsection{Inference}
For brevity, use ${\boldsymbol{\theta}}$ to represent the parameters of the distribution ($\Pwd$, $\qB$, $\qL$, $\qH$, $\alpha$, and $K_{\rm th}$). In addition, the distribution depends on each binary's primary mass and orbital period through the detection function.
To simplify the inference, these values are fixed at nominal literature values. The likelihood of the $i$\textsuperscript{th} spectroscopic binary is given by
\begin{equation} 
{\mathcal{L}_i\big(\boldsymbol{\theta}\,|\, S_i\big)} = f_S^{\rm obs}\big(S_i\,|\, \boldsymbol{\theta}, \, {m_{\scriptsize\textsc{rg},i}},\, P_i\big).
\end{equation} 
The term on the right-hand side represents the modified mass function distribution corresponding to the mass ratio distribution from equation~(\ref{eq: fq_obs}), derived according to the prescription in Appendix~\ref{app: mmf dist}. The equation above does not account for the measurement uncertainty in the modified mass function. While this compromise may be acceptable in some cases, a more general consideration may be required.\footnote{The reduced mass function ($y$) is a ratio of two quantities. Assuming measurement errors are Gaussian distributed, the normal ratio distribution (e.g., \citealt{Hinkley1969}) can provide the uncertainty estimates for $y$, which can be incorporated into the likelihood.}

%
\begin{figure}
	\centering
	\includegraphics[width=0.975\linewidth ,trim={0 0 0 0 },clip]{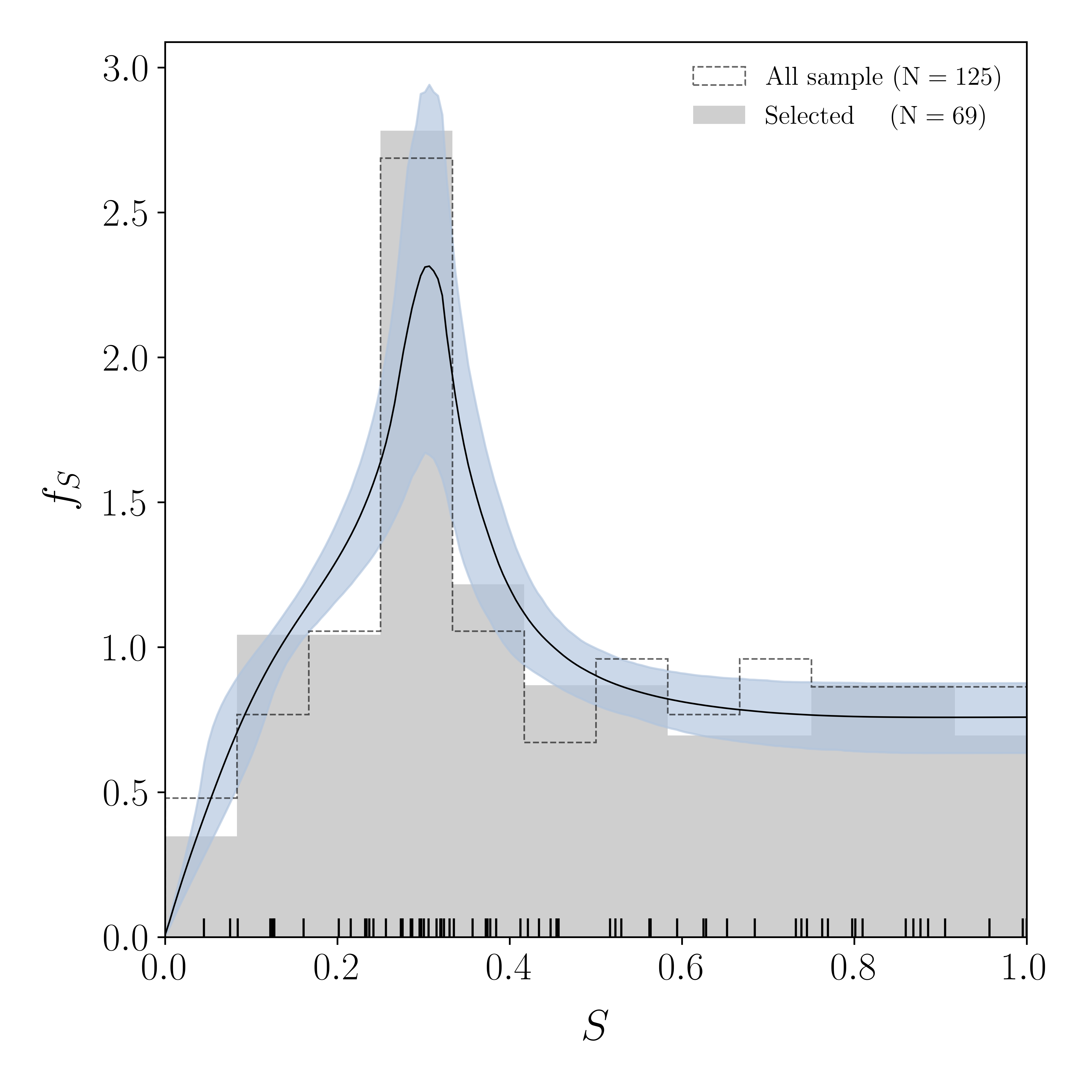}
	\caption{A histogram showing the modified mass function values of the sample. Grey bars show the distribution selected subsample of $69$ systems and the entire sample of $125$ binaries is overlaid as a black dashed line. The black ticks at the bottom of the figure indicate the position of individual targets in the selected samples.  The solid black line shows the expected shape of the distribution, using the fitted parameters from Section~\ref{sec: merm07 sample} and after accounting for the radial velocity detection threshold, $\mathcal{D}$. The light blue area represents the corresponding one-sigma confidence interval.}
	\label{fig:merm07_mmfhist}
\end{figure}

The parameters are determined by sampling the logarithm of the posterior distribution:
\begin{equation}
    \log{P}\big(\boldsymbol{\theta}|\{S_i\}\big) = \sum_i\log{\mathcal{L}_i}\big(\boldsymbol{\theta}\,|\, S_i\big) + \log\pi\big(\boldsymbol{\theta}\big),
\end{equation}
where $\pi$ represents the prior distribution of the fitted parameters. A normal prior is applied to $\qH$ (mean $0.58$, standard deviation $0.10$), while flat priors are used for $\Pwd$, $\qB$, and $\qL$ within $0 < \Pwd < 1$ and $0 < \qB < \qL < \qH < 1$. The prior for $K_{\rm th}$ is flat between $0$ and $5$~km~s\textsuperscript{-1}, and  $\alpha$ is assumed to be normally distributed around $2.35$, with a standard deviation of $0.30$.

\begin{figure}
	\centering
	\includegraphics[width=0.99\linewidth ,trim={0 0 0 0 },clip]{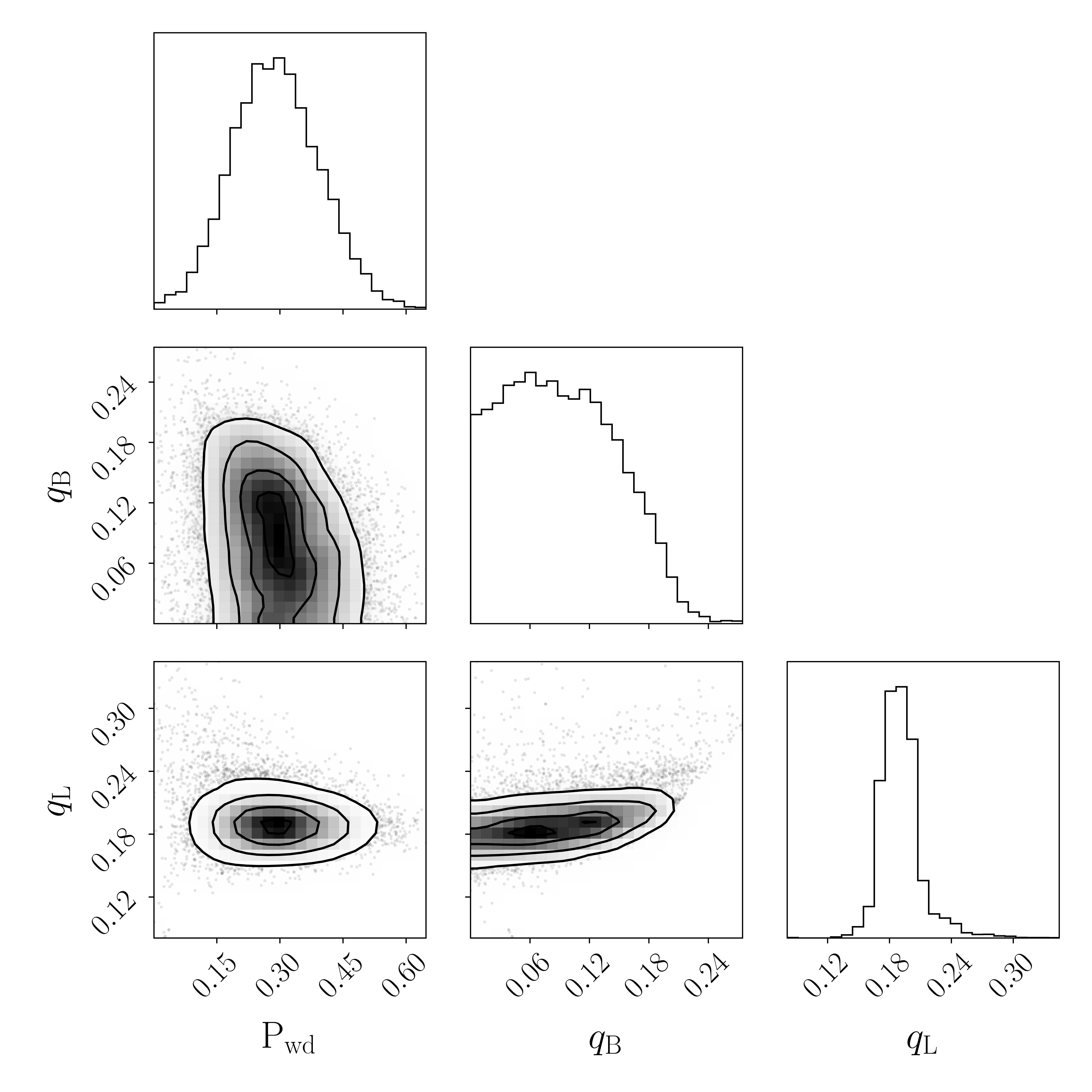}
	\caption{A corner plot depicting the posterior distribution of the first three fitted parameters. The white dwarf fraction, $\Pwd$, and mass ratio cutoff value, $\qL$, are well constrained and consistent with the initial qualitative estimates. However, the data is more limited in constraining $\qB$ and sets a firm upper limit at ${\sim}\,0.2$, with all values below ${\sim}\,0.15$ equally probable. Two parameters were excluded from the plot, $\qH$ and $K_{\rm th}$. The full corner plot is available in the online supplementary material.} 
	\label{fig:merm07_posterior}
\end{figure}

The analysis is performed on the 69 selected systems described above. To estimate the parameters, we used \texttt{emcee} \citep{foreman-Mackey2013}, with 50 walkers taking 5000 steps each. Figures~\ref{fig:merm07_mmfhist} and~\ref{fig:merm07_posterior} show the estimated $f_S$ distribution and posterior corner plots. Evidently, the analysis provides tighter constraints on $\qL$ than on $\qB$, as expected for heavy-tailed distributions. Cutoff values, such as $\qL$, are frequently sampled, while tail values, determining $\qB$, are rare. Consequently, $\qB$ is constrained to an upper limit of ${\sim}\,0.2$, with values below ${\sim}\,0.15$ being roughly equally probable. The estimated parameter values, based on the 16\textsuperscript{th}, 50\textsuperscript{th}, and 84\textsuperscript{th} posterior percentiles, are:
\begin{equation}
    \begin{aligned}
        &\Pwd &= 0.29_{-0.11}^{+0.10},\,\,\,\,\,&
        &\qB  &= 0.088_{-0.057}^{+0.064},&\\[8pt]
        &\qL  &= 0.189_{-0.015}^{+0.017},&
        &\qH  &= 0.58_{-0.10}^{+0.10}.&
    \end{aligned}
\end{equation}
The estimated detection threshold for radial velocity, $K_{\rm th}$, is $1.1 \pm 0.5 ~\text{km}~\text{s}^{-1}$. The sample does not constrain the initial mass function, yielding $\alpha = 2.36 \pm 0.29$. 

Using posterior samples of the mass ratio distribution, we estimated the white dwarf mass distribution in our sample through a Monte Carlo experiment. For each posterior step in the \texttt{emcee} chain, a primary mass was randomly assigned, assuming $\mRG/\msun \sim \mathcal{N}(2.4, 0.4^2)$. The white dwarf mass distribution was computed on a predefined mass grid based on the posterior samples, the assigned giant mass, and equation~(\ref{eq:qBounds}). Repeating this procedure generated an ensemble of $\mWD$ distributions. The distribution shown in Figure~\ref{fig:mWD_posterior} was obtained by averaging the values at each grid point. The blue-shaded region spans the 16\textsuperscript{th} to 84\textsuperscript{th} percentiles ($0.43$–$0.73~\msun$). Vertical lines indicate the distribution's mode, median, and mean ($0.50$, $0.54$, and $0.58~\msun$, respectively). 
The mean mass estimate is consistent with the results reported by \citet{north14} and \citet{swaelmen17}. However, the analysis relies on the heavy-tailed Pareto distribution, suggesting that most white dwarfs are less massive than this mean value. Interestingly, the white dwarfs in this sample are less massive than expected, given their expected progenitor masses. This is discussed further below.

Now, we use the fitted parameters of the distribution to constrain the IFMR: the parameters of equation~(\ref{EQ: IFMR main}) were derived from the posterior samples, assuming the red giant's current mass closely approximates its zero-age main-sequence mass. For each posterior sample, a primary mass was drawn as described earlier, and a random mass gap was assigned, where $\Delta/\msun \sim {\rm Uniform}(0.2, 0.4)$; see the discussion in Section~\ref{sec: popsyn}. The turn-off mass was then calculated using equation~(\ref{eq:Owd}), with samples retained only if $m_\tau < \mRG < m_\tau + \Delta$. The IFMR slope and intercept were determined using equations~(\ref{EQ: IFMR main}) and~(\ref{eq:qBounds}). Repeating this process yielded a distribution of IFMR parameters. 

Considering the sample size and the estimated value of $\Pwd$,  we expect to have $20\pm8$ white dwarfs in the observed population. This small sample and the strong correlations between the parameters limit the precision of the inferred IFMR. The derived relation is 
\begin{equation}
    \label{eq: ABfit}
    \mWD \simeq \big(0.097 \pm 0.050 \big) \, \mPR + \big(0.19 \pm 0.15)\, \,\,\msun,
\end{equation}
where the parameters exhibit a strong linear dependence with a correlation coefficient of $-0.89$. Despite these limitations, valuable insights can still be obtained. Figure~\ref{fig:fqpeak} provides a contour map of the IFMR parameters, showing 1–2$\sigma$ confidence levels.

The derived distribution of the parameters aligns with the relation reported by \citet{Cunningham2024} for progenitors more massive than ${\sim}\,5~\msun$, which is close to a scaling relation. Therefore, to get a better qualitative estimate of the relation, we estimate the IFMR for our sample again using only samples of negligible intercept ($\qB < 1\%$), yielding
\begin{equation}
    \label{eq: A0fit}
    A_0\equiv\frac{\mWD}{\mPR}\approx 17.0 \pm 1.3 \%.
\end{equation}
This approximated relation is further discussed below.

%

\section{Discussion}
\label{sec: discussion}

\subsection{Evidence for enhanced mass loss}
\begin{figure}
	\centering
	\includegraphics[width=0.99\linewidth ,trim={0 0 0 0 },clip]{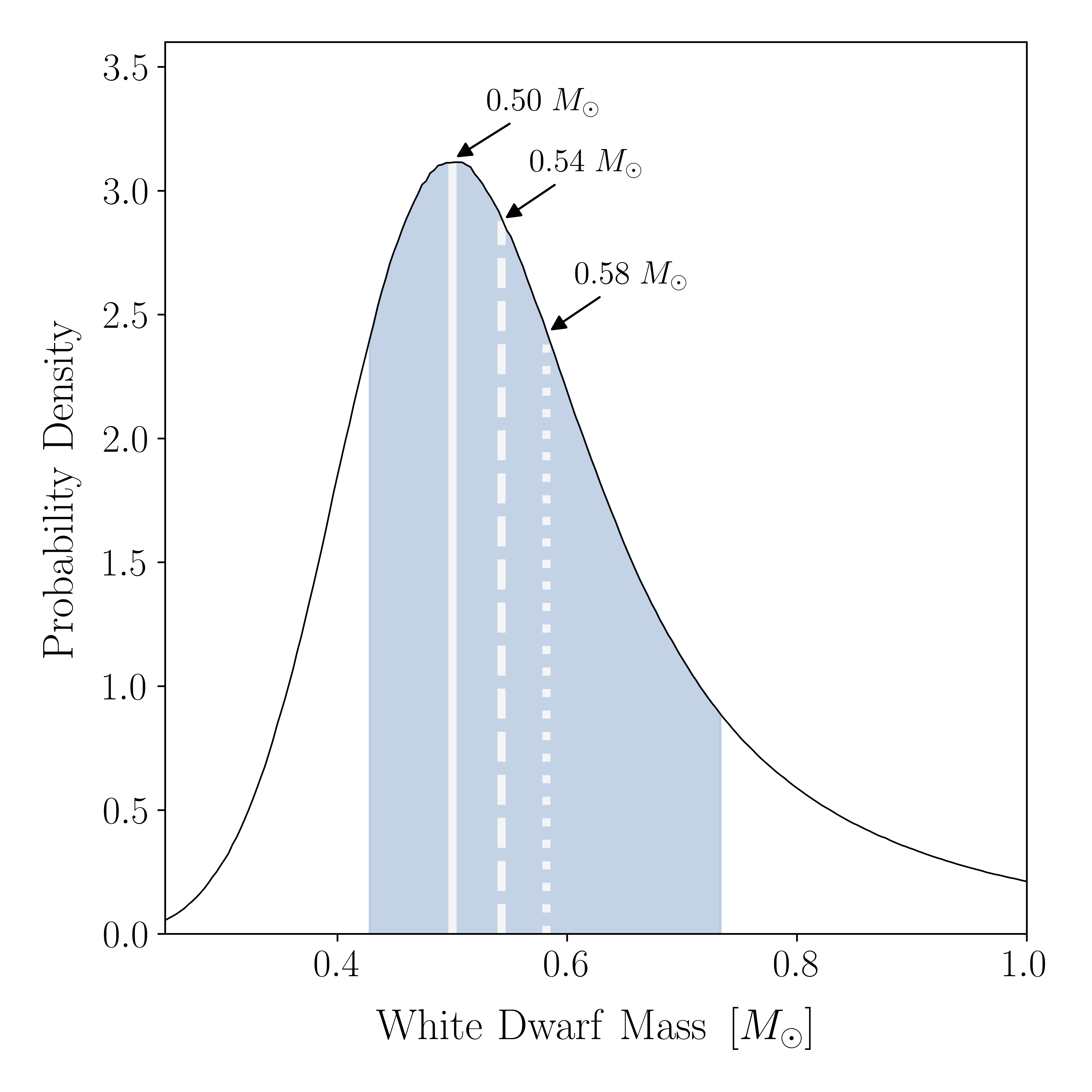}
	\caption{The derived white dwarf mass distribution, based on analyzing the selected sample of 69 binary systems, out of which $20\pm8$ systems comprise a red giant with a white dwarf companion. The solid black line depicts the estimated mass probability density function. The blue-shaded area covers the 16\textsuperscript{th} to 84\textsuperscript{th} percentile range ($0.43-0.73~\msun$). The solid, dashed, and dotted white lines illustrate the position of the distribution's mode, median, and mean.} 
	\label{fig:mWD_posterior}
\end{figure}

The cutoff value at $\qL \simeq 19\%$ indicates that half of the white dwarfs in this sample have masses below ${\sim}\,0.55~\msun$, based on the median value of the corresponding Pareto distribution. The mass distribution in Figure~\ref{fig:mWD_posterior} supports this estimate. This result suggests that the white dwarfs in this sample are generally less massive than their field counterparts, which are sharply centered around $0.6~\msun$, with only ${\sim}\,10\%$ having masses below ${\sim}\,0.55~\msun$ (e.g., \citealt{Kilic2020, Jimenez-Esteban2023, O'Brien2024}).

The discrepancy is more pronounced given that the progenitors in this sample likely exceeded ${\sim}\,2.5~\msun$. Current relations predict such progenitors would produce remnants more massive than ${\sim}\,0.65~\msun$ (e.g., \citealt{cummings16, El-Badry2018, Cunningham2024, Hollands2024}, but see \citealt{Andrews2015, Choi2016}). These white dwarfs are, therefore, $10-20\%$ less massive than expected from single-star evolution. A caveat of this argument is the possibility of a significant mass transfer phase, implying that the giants -- and consequently the white dwarf progenitors -- could have initially been less massive. However, the cluster’s age estimate of ${\sim}\,1$~Gyr mitigates this concern: with turn-off masses near $2~\msun$ (\citealt{swaelmen17}), the mass discrepancy remains.

The low masses of these white dwarfs likely reflect the impact of binary interactions. For instance, extremely low-mass white dwarfs ($\lesssim 0.3~\msun$) are commonly found in close binary systems with orbital periods of hours to days \citep[e.g.,]{Li2019, Brown2010}. This highlights how binarity can significantly alter stellar evolution, leading to remnants with lower masses. In comparison, the binaries in \citet{merm07cat} host more massive white dwarfs in systems with significantly wider separations. Thus, the systems in this study likely experienced milder, but still non-negligible, interaction.

With turnoff masses of $2$–$2.5~\msun$ and using the \citet{salpeter55} initial mass function, the median progenitor mass for the white dwarfs in this sample is estimated to be $3$–$3.5~\msun$. This narrow range justifies applying the scaling relation in equation~(\ref{eq: A0fit}) instead of the linear model in equation~(\ref{eq: ABfit}). The scaling relation indicates that a typical progenitor in this sample lost $80$–$85\%$ of its mass while evolving into a white dwarf. While this level of mass loss is consistent with expectations for stars more massive than ${\sim}\,5~\msun$ \citep{cummings18}, the initial mass function suggests that at least $80\%$ of the progenitors had initial masses below ${\sim}\,5~\msun$, where the expected mass loss fraction is closer to $75\%$. This implies an additional ${\sim}\,0.2~\msun$ of mass loss, likely due to binary interactions.

Still, while this analysis estimates the amount of mass loss, it does not determine how or during which evolutionary phase the mass was lost. Nonetheless, despite the modest sample size, the analysis effectively constrains the properties of the population. 

%

\subsection{Assumptions and limitations}
The analysis above relies on several assumptions. First, the primordial mass ratio distribution is assumed to be flat. Admittedly, this may be an oversimplification, as evidenced by conflicting results in the literature. While some studies suggest that the mass ratio distribution of spectroscopic binaries is indeed flat (\citealt{mazeh03, raghavan10}), others report power-law behavior or a significant population of twin binaries (e.g., \citealt{El-Badry2019}).

The impact of a population of twins is relatively straightforward to address. Equal-mass binaries leave the main sequence roughly simultaneously, making it possible to identify both components in some cases. Moreover, twins produce a distinct peak in the mass ratio distribution, which is well-separated from the white dwarf Pareto peak for giants more massive than ${\sim}\,1\,\msun$. This separation enables their inclusion in the analysis by explicitly accounting for their contribution or parameterizing it as needed. A non-uniform mass ratio distribution poses a more significant challenge, as it affects both the estimated fraction of white dwarfs and the shape of their mass distribution.
Nevertheless, assuming a flat distribution provides a reasonable first approximation, given the current sample size. Most conclusions in this work rely on the position of the distribution's peak, corresponding to the least massive progenitor capable of evolving into a white dwarf. This key property is robust against variations in the underlying mass ratio distribution. The model can be generalized further to account for non-uniform distribution if needed. 

The assumed shape of the initial mass function may also influence the results. This consideration is particularly relevant given the possibility that it is not universal. The exponent, $\alpha$, affects the white dwarf fraction and the slope of the mass distribution but not its cutoff value. The prior used in this analysis is consistent within roughly $1\sigma$ with the initial mass function exponent found for the thin disk, thick disk, and high-metallicity halo by \citet[][$2.05 \lesssim \alpha \lesssim 2.75$]{hallakoun21}. However, the sample analyzed in this work is not large enough to constrain $\alpha$. Nevertheless, when considering larger samples or populations obtained from distinct Galactic environments, the assumptions regarding the initial mass function could become more significant.

Another limitation of the method presented here stems from the inherent covariance between the parameters of the IFMR. These parameters are related to one another through the mass ratio cutoff value, $\qL$, as equation~(\ref{eq:qBounds}) shows. Strong constraints on $\qB$ are required to mitigate this degeneracy, but these are difficult to achieve with small sample sizes, as they depend on accurate estimates of the distribution slope. Nevertheless, by leveraging simple supporting arguments, the analysis above offers meaningful insights into the history of mass loss in these systems.

\section{Summary and conclusions}
\label{sec: summary}
This study demonstrates the utility of mass ratio distributions as a tool to probe the influence of binarity on stellar evolution. The truncated Pareto profile used to describe the population provides effective constraints on the properties of white dwarfs and their progenitors. Analysis of the \citet{mermilliod07} sample reveals that, in the explored regime of orbital separations, white dwarfs are ${\sim}\,20\%$ less massive than their isolated counterparts. The findings suggest that their progenitors experienced enhanced mass loss of approximately ${\sim}\,0.2~\msun$ due to binary interactions.

Mass transfer and binary co-evolution in this regime, on the cusp of a common envelope phase, is not fully understood. Investigating the relationship between white dwarf mass and binary system properties will help constrain the dominant mechanisms and quantify their role in shaping the population. Expanding the analysis to larger samples will enable the search for dependence between the white dwarf properties and the system's physical parameters, such as orbital separation, eccentricity, and chemical composition. Such a dependency, if detected, could yield insights into the final stages of stellar evolution \citep[e.g.,][]{Shahaf2023, Shahaf2024, Hallakoun2024, Yamaguchi2024, Yamaguchi2024b}. 

With larger samples, further work can also focus on stellar core growth during the giant and asymptotic giant branch phases. The \textit{Gaia} third data release has already provided valuable opportunities for this purpose \citep[e.g., using the][catalog]{Hunt2023}. These samples can help constrain the IFMR for binaries with intermediate separations (Ironi et al., in preparation). Moreover, additional observations, such as measurements of Barium enhancement, can reveal whether the white dwarf progenitor underwent an asymptotic giant phase while providing independent constraints on its mass \citep[e.g.,][]{Rekhi2024}. Such constraints can, in turn, identify when core growth is interrupted and statistically evaluate its impact on the remnant's mass. Observations of this kind can probe the mass accumulation rate during the asymptotic giant branch phase \citep{Marigo2022, Marigo2022b, Addari2024}.

Another promising avenue to increase sample sizes is to relax the requirement for red giants to belong to open clusters. For example, chemical abundance indicators, such as carbon-nitrogen ratios, can help identify suitable red giant systems for study (e.g., \citealt{Roberts2024}; but see the limitations discussed by \citealt{Bufanda2023}). Extending the method to binaries with main-sequence companions could also be viable; since the red giant’s evolutionary state does not directly affect the mass ratio distribution, such systems could offer complementary insights. However, potential biases introduced by age heterogeneity in main-sequence systems must be carefully mitigated to ensure reliable results. Data from Gaia and next-generation surveys will be instrumental in refining this framework and uncovering new details about the final stages of stellar evolution.

\begin{acknowledgments}
I thank the anonymous referee for the helpful comments and suggestions. 
I thank Tsevi Mazeh and Simchon Faigler for their contributions during the early stages of this work, and I am grateful to Na'ama Hallakoun for her invaluable advice and support. I also thank Sagi Ben-Ami, Dotan Gazith, Oryna Ivashtenko, Dan Maoz, Eran Ofek, Hans-Walter Rix, Tomer Shenar, and Barak Zackay for their comments and feedback. SS is supported by a Benoziyo Prize Postdoctoral Fellowship. This work is dedicated to the memory of Daniel Alloush and Tom Ish-Shalom; you will not be forgotten. 
\end{acknowledgments}

%
\software{This work used \texttt{cogsworth} and its dependencies \citep{Wagg2024, Wagg2025, Breivik2020}; the Unified Cluster Catalogue \citep{Perren2023}; \texttt{emcee} \citep{foreman-Mackey2013}; \texttt{numpy} \citep{Numpy_2006, Numpy_2011}; and \texttt{scipy} \citep{Virtanen_2020}.} 
%


\appendix
\section{The reduced and modified mass functions distribution}
\label{app: mmf dist}
For a given mass ratio distribution, the reduced mass function distribution can be obtained by assuming that the orbital plane of motion is randomly oriented (e.g., \citealt{heacox95}, \citealt{shahaf17}). The distribution is expressed in an integral form, 
\begin{equation}
    \label{eq:fy}
    f_y(y; f_q) = {\int_{\qmin}^1 {f_q(q) \, \mathbb{K}\big[y,q\big] \, dq }}\, ,
\end{equation}
where  
%
\begin{equation*}
\mathbb{K}(y,q) 
= \frac{(1+q)^{4/3}}{ 3 \, y^{1/3} \, q \,  \sqrt{ q^2 - y^{2/3}(1+q)^{4/3} }  }  \,.
\end{equation*}	
%
The modified mass function distribution, $f_S$, is related to that of the reduced mass function via
\begin{equation}
    \label{eq:fS}
    f_S(S; f_q) = \frac{f_y\big[y(S); f_q\big]}{f_y\big[y(S); 1\big]}\, 
\end{equation}
where $y(S)$ is the inverse function of equation~(\ref{EQ:S}). The distribution of $S$ mimics the functional shape of the unknown mass ratio distribution, as demonstrated in Figure~\ref{fig:fS}. For additional examples of using the modified mass function as a proxy for the mass ratio distribution, see \citet{Shahaf2019}.

\bibliography{main}{}

\begin{thebibliography}{}
\expandafter\ifx\csname natexlab\endcsname\relax\def\natexlab#1{#1}\fi
\providecommand{\url}[1]{\href{#1}{#1}}
\providecommand{\dodoi}[1]{doi:~\href{http://doi.org/#1}{\nolinkurl{#1}}}
\providecommand{\doeprint}[1]{\href{http://ascl.net/#1}{\nolinkurl{http://ascl.net/#1}}}
\providecommand{\doarXiv}[1]{\href{https://arxiv.org/abs/#1}{\nolinkurl{https://arxiv.org/abs/#1}}}

\bibitem[{{Addari} {et~al.}(2024){Addari}, {Marigo}, {Bressan}, {Costa},
  {Shepherd}, \& {Volpato}}]{Addari2024}
{Addari}, F., {Marigo}, P., {Bressan}, A., {et~al.} 2024, \apj, 964, 51,
  \dodoi{10.3847/1538-4357/ad2067}

\bibitem[{{Althaus} {et~al.}(2010){Althaus}, {C{\'o}rsico}, {Isern}, \&
  {Garc{\'\i}a-Berro}}]{Althaus2010}
{Althaus}, L.~G., {C{\'o}rsico}, A.~H., {Isern}, J., \& {Garc{\'\i}a-Berro}, E.
  2010, \aapr, 18, 471, \dodoi{10.1007/s00159-010-0033-1}

\bibitem[{{Andrews} {et~al.}(2015){Andrews}, {Ag{\"u}eros}, {Gianninas},
  {Kilic}, {Dhital}, \& {Anderson}}]{Andrews2015}
{Andrews}, J.~J., {Ag{\"u}eros}, M.~A., {Gianninas}, A., {et~al.} 2015, \apj,
  815, 63, \dodoi{10.1088/0004-637X/815/1/63}

\bibitem[{{Barrientos} \& {Chanam{\'e}}(2021)}]{Barrientos2021}
{Barrientos}, M., \& {Chanam{\'e}}, J. 2021, \apj, 923, 181,
  \dodoi{10.3847/1538-4357/ac2f49}

\bibitem[{{Beuther} {et~al.}(2019){Beuther}, {Ahmadi}, {Mottram}, {Linz},
  {Maud}, {Henning}, {Kuiper}, {Walsh}, {Johnston}, \& {Longmore}}]{beuther19}
{Beuther}, H., {Ahmadi}, A., {Mottram}, J.~C., {et~al.} 2019, \aap, 621, A122,
  \dodoi{10.1051/0004-6361/201834064}

\bibitem[{{Boffin} {et~al.}(1993){Boffin}, {Cerf}, \& {Paulus}}]{boffin93}
{Boffin}, H.~M.~J., {Cerf}, N., \& {Paulus}, G. 1993, \aap, 271, 125

\bibitem[{{Breivik} {et~al.}(2020){Breivik}, {Coughlin}, {Zevin}, {Rodriguez},
  {Kremer}, \& et~al.}]{Breivik2020}
{Breivik}, K., {Coughlin}, S., {Zevin}, M., {et~al.} 2020, \apj, 898, 71,
  \dodoi{10.3847/1538-4357/ab9d85}

\bibitem[{{Bressan} {et~al.}(2012){Bressan}, {Marigo}, {Girardi}, {Salasnich},
  {Dal Cero}, {Rubele}, \& {Nanni}}]{Bressan_2012}
{Bressan}, A., {Marigo}, P., {Girardi}, L., {et~al.} 2012, \mnras, 427, 127,
  \dodoi{10.1111/j.1365-2966.2012.21948.x}

\bibitem[{{Brown} {et~al.}(2010){Brown}, {Kilic}, {Allende Prieto}, \&
  {Kenyon}}]{Brown2010}
{Brown}, W.~R., {Kilic}, M., {Allende Prieto}, C., \& {Kenyon}, S.~J. 2010,
  \apj, 723, 1072, \dodoi{10.1088/0004-637X/723/2/1072}

\bibitem[{{Bufanda} {et~al.}(2023){Bufanda}, {Tayar}, {Huber}, {Hasselquist},
  \& {Lane}}]{Bufanda2023}
{Bufanda}, E., {Tayar}, J., {Huber}, D., {Hasselquist}, S., \& {Lane}, R.~R.
  2023, \apj, 959, 123, \dodoi{10.3847/1538-4357/acf9a5}

\bibitem[{{Chen} {et~al.}(2015){Chen}, {Bressan}, {Girardi}, {Marigo}, {Kong},
  \& {Lanza}}]{Chen_2015}
{Chen}, Y., {Bressan}, A., {Girardi}, L., {et~al.} 2015, \mnras, 452, 1068,
  \dodoi{10.1093/mnras/stv1281}

\bibitem[{{Chen} {et~al.}(2014){Chen}, {Girardi}, {Bressan}, {Marigo},
  {Barbieri}, \& {Kong}}]{Chen_2014}
{Chen}, Y., {Girardi}, L., {Bressan}, A., {et~al.} 2014, \mnras, 444, 2525,
  \dodoi{10.1093/mnras/stu1605}

\bibitem[{{Cheng} {et~al.}(2020){Cheng}, {Cummings}, {M{\'e}nard}, \&
  {Toonen}}]{Cheng2020}
{Cheng}, S., {Cummings}, J.~D., {M{\'e}nard}, B., \& {Toonen}, S. 2020, \apj,
  891, 160, \dodoi{10.3847/1538-4357/ab733c}

\bibitem[{{Choi} {et~al.}(2016){Choi}, {Dotter}, {Conroy}, {Cantiello},
  {Paxton}, \& {Johnson}}]{Choi2016}
{Choi}, J., {Dotter}, A., {Conroy}, C., {et~al.} 2016, \apj, 823, 102,
  \dodoi{10.3847/0004-637X/823/2/102}

\bibitem[{{Cummings} {et~al.}(2016){Cummings}, {Kalirai}, {Tremblay}, \&
  {Ramirez-Ruiz}}]{cummings16}
{Cummings}, J.~D., {Kalirai}, J.~S., {Tremblay}, P.-E., \& {Ramirez-Ruiz}, E.
  2016, \apj, 818, 84, \dodoi{10.3847/0004-637X/818/1/84}

\bibitem[{{Cummings} {et~al.}(2018){Cummings}, {Kalirai}, {Tremblay},
  {Ramirez-Ruiz}, \& {Choi}}]{cummings18}
{Cummings}, J.~D., {Kalirai}, J.~S., {Tremblay}, P.~E., {Ramirez-Ruiz}, E., \&
  {Choi}, J. 2018, \apj, 866, 21, \dodoi{10.3847/1538-4357/aadfd6}

\bibitem[{{Cunningham} {et~al.}(2024){Cunningham}, {Tremblay}, \& {W.
  O'Brien}}]{Cunningham2024}
{Cunningham}, T., {Tremblay}, P.-E., \& {W. O'Brien}, M. 2024, \mnras, 527,
  3602, \dodoi{10.1093/mnras/stad3275}

\bibitem[{{Dias} {et~al.}(2021){Dias}, {Monteiro}, {Moitinho}, {L{\'e}pine},
  {Carraro}, {Paunzen}, {Alessi}, \& {Villela}}]{Dias2021}
{Dias}, W.~S., {Monteiro}, H., {Moitinho}, A., {et~al.} 2021, \mnras, 504, 356,
  \dodoi{10.1093/mnras/stab770}

\bibitem[{{Duch{\^e}ne} \& {Kraus}(2013)}]{Duchene2013}
{Duch{\^e}ne}, G., \& {Kraus}, A. 2013, \araa, 51, 269,
  \dodoi{10.1146/annurev-astro-081710-102602}

\bibitem[{{El-Badry}(2024)}]{El-Badry2024}
{El-Badry}, K. 2024, \nar, 98, 101694, \dodoi{10.1016/j.newar.2024.101694}

\bibitem[{{El-Badry} {et~al.}(2019){El-Badry}, {Rix}, {Tian}, {Duch{\^e}ne}, \&
  {Moe}}]{El-Badry2019}
{El-Badry}, K., {Rix}, H.-W., {Tian}, H., {Duch{\^e}ne}, G., \& {Moe}, M. 2019,
  \mnras, 489, 5822, \dodoi{10.1093/mnras/stz2480}

\bibitem[{{El-Badry} {et~al.}(2018){El-Badry}, {Rix}, \&
  {Weisz}}]{El-Badry2018}
{El-Badry}, K., {Rix}, H.-W., \& {Weisz}, D.~R. 2018, \apjl, 860, L17,
  \dodoi{10.3847/2041-8213/aaca9c}

\bibitem[{{Fleury} {et~al.}(2022){Fleury}, {Caiazzo}, \& {Heyl}}]{Fleury2022}
{Fleury}, L., {Caiazzo}, I., \& {Heyl}, J. 2022, \mnras, 511, 5984,
  \dodoi{10.1093/mnras/stac458}

\bibitem[{{Foreman-Mackey} {et~al.}(2013){Foreman-Mackey}, {Conley},
  {Meierjurgen Farr}, {Hogg}, {Lang}, {Marshall}, {Price-Whelan}, {Sanders}, \&
  {Zuntz}}]{foreman-Mackey2013}
{Foreman-Mackey}, D., {Conley}, A., {Meierjurgen Farr}, W., {et~al.} 2013,
  {emcee: The MCMC Hammer}, Astrophysics Source Code Library.
\newblock \doeprint{1303.002}

\bibitem[{{Gentile Fusillo} {et~al.}(2021){Gentile Fusillo}, {Tremblay},
  {Cukanovaite}, {Vorontseva}, {Lallement}, {Hollands}, {G{\"a}nsicke},
  {Burdge}, {McCleery}, \& {Jordan}}]{GentileFusillo2021}
{Gentile Fusillo}, N.~P., {Tremblay}, P.~E., {Cukanovaite}, E., {et~al.} 2021,
  \mnras, 508, 3877, \dodoi{10.1093/mnras/stab2672}

\bibitem[{{Hallakoun} \& {Maoz}(2021)}]{hallakoun21}
{Hallakoun}, N., \& {Maoz}, D. 2021, \mnras, 507, 398,
  \dodoi{10.1093/mnras/stab2145}

\bibitem[{{Hallakoun} {et~al.}(2024){Hallakoun}, {Shahaf}, {Mazeh}, {Toonen},
  \& {Ben-Ami}}]{Hallakoun2024}
{Hallakoun}, N., {Shahaf}, S., {Mazeh}, T., {Toonen}, S., \& {Ben-Ami}, S.
  2024, \apjl, 970, L11, \dodoi{10.3847/2041-8213/ad5e63}

\bibitem[{{Han} {et~al.}(1995){Han}, {Podsiadlowski}, \& {Eggleton}}]{Han1995}
{Han}, Z., {Podsiadlowski}, P., \& {Eggleton}, P.~P. 1995, \mnras, 272, 800,
  \dodoi{10.1093/mnras/272.4.800}

\bibitem[{{Heacox}(1995)}]{heacox95}
{Heacox}, W.~D. 1995, \aj, 109, 2670, \dodoi{10.1086/117480}

\bibitem[{{Heintz} {et~al.}(2022){Heintz}, {Hermes}, {El-Badry}, {Walsh}, {van
  Saders}, {Fields}, \& {Koester}}]{Heintz2022}
{Heintz}, T.~M., {Hermes}, J.~J., {El-Badry}, K., {et~al.} 2022, \apj, 934,
  148, \dodoi{10.3847/1538-4357/ac78d9}

\bibitem[{Hinkley(1969)}]{Hinkley1969}
Hinkley, D.~V. 1969, Biometrika, 56, 635, \dodoi{10.1093/biomet/56.3.635}

\bibitem[{{Holberg} {et~al.}(2013){Holberg}, {Oswalt}, {Sion}, {Barstow}, \&
  {Burleigh}}]{Holberg2013}
{Holberg}, J.~B., {Oswalt}, T.~D., {Sion}, E.~M., {Barstow}, M.~A., \&
  {Burleigh}, M.~R. 2013, \mnras, 435, 2077, \dodoi{10.1093/mnras/stt1433}

\bibitem[{{Hollands} {et~al.}(2024){Hollands}, {Littlefair}, \&
  {Parsons}}]{Hollands2024}
{Hollands}, M.~A., {Littlefair}, S.~P., \& {Parsons}, S.~G. 2024, \mnras, 527,
  9061, \dodoi{10.1093/mnras/stad3729}

\bibitem[{{Hunt} \& {Reffert}(2023)}]{Hunt2023}
{Hunt}, E.~L., \& {Reffert}, S. 2023, \aap, 673, A114,
  \dodoi{10.1051/0004-6361/202346285}

\bibitem[{{Isern} {et~al.}(2022){Isern}, {Torres}, \&
  {Rebassa-Mansergas}}]{Isern2022}
{Isern}, J., {Torres}, S., \& {Rebassa-Mansergas}, A. 2022, Frontiers in
  Astronomy and Space Sciences, 9, 6, \dodoi{10.3389/fspas.2022.815517}

\bibitem[{{Jim{\'e}nez-Esteban} {et~al.}(2023){Jim{\'e}nez-Esteban}, {Torres},
  {Rebassa-Mansergas}, {Cruz}, {Murillo-Ojeda}, {Solano}, {Rodrigo}, \&
  {Camisassa}}]{Jimenez-Esteban2023}
{Jim{\'e}nez-Esteban}, F.~M., {Torres}, S., {Rebassa-Mansergas}, A., {et~al.}
  2023, \mnras, 518, 5106, \dodoi{10.1093/mnras/stac3382}

\bibitem[{{Kilic} {et~al.}(2020){Kilic}, {Bergeron}, {Kosakowski}, {Brown},
  {Ag{\"u}eros}, \& {Blouin}}]{Kilic2020}
{Kilic}, M., {Bergeron}, P., {Kosakowski}, A., {et~al.} 2020, \apj, 898, 84,
  \dodoi{10.3847/1538-4357/ab9b8d}

\bibitem[{{Kilic} {et~al.}(2023){Kilic}, {Moss}, {Kosakowski}, {Bergeron},
  {Conly}, {Brown}, {Toonen}, {Williams}, \& {Dufour}}]{Kilic2023}
{Kilic}, M., {Moss}, A.~G., {Kosakowski}, A., {et~al.} 2023, \mnras, 518, 2341,
  \dodoi{10.1093/mnras/stac3182}

\bibitem[{{Kosakowski} {et~al.}(2023){Kosakowski}, {Brown}, {Kilic}, {Kupfer},
  {B{\'e}dard}, {Gianninas}, {Ag{\"u}eros}, \& {Barrientos}}]{Kosakowski2023}
{Kosakowski}, A., {Brown}, W.~R., {Kilic}, M., {et~al.} 2023, \apj, 950, 141,
  \dodoi{10.3847/1538-4357/acd187}

\bibitem[{{Kuiper}(1935)}]{kuiper35}
{Kuiper}, G.~P. 1935, \pasp, 47, 15, \dodoi{10.1086/124531}

\bibitem[{{Larson}(1972)}]{larson72}
{Larson}, R.~B. 1972, \mnras, 156, 437, \dodoi{10.1093/mnras/156.4.437}

\bibitem[{{Li} {et~al.}(2019){Li}, {Chen}, {Chen}, \& {Han}}]{Li2019}
{Li}, Z., {Chen}, X., {Chen}, H.-L., \& {Han}, Z. 2019, \apj, 871, 148,
  \dodoi{10.3847/1538-4357/aaf9a1}

\bibitem[{{Marigo}(2022)}]{Marigo2022b}
{Marigo}, P. 2022, Universe, 8, 243, \dodoi{10.3390/universe8040243}

\bibitem[{{Marigo} {et~al.}(2022){Marigo}, {Bossini}, {Trabucchi}, {Addari},
  {Girardi}, {Cummings}, {Pastorelli}, {Dal Tio}, {Costa}, \&
  {Bressan}}]{Marigo2022}
{Marigo}, P., {Bossini}, D., {Trabucchi}, M., {et~al.} 2022, \apjs, 258, 43,
  \dodoi{10.3847/1538-4365/ac374a}

\bibitem[{{Mazeh} \& {Goldberg}(1992)}]{mg92}
{Mazeh}, T., \& {Goldberg}, D. 1992, \apj, 394, 592, \dodoi{10.1086/171611}

\bibitem[{{Mazeh} {et~al.}(2003){Mazeh}, {Simon}, {Prato}, {Markus}, \&
  {Zucker}}]{mazeh03}
{Mazeh}, T., {Simon}, M., {Prato}, L., {Markus}, B., \& {Zucker}, S. 2003,
  \apj, 599, 1344, \dodoi{10.1086/379346}

\bibitem[{{Mermilliod} {et~al.}(2007{\natexlab{a}}){Mermilliod}, {Andersen},
  {Latham}, \& {Mayor}}]{merm07cat}
{Mermilliod}, J.~C., {Andersen}, J., {Latham}, D., \& {Mayor}, M.
  2007{\natexlab{a}}, {VizieR Online Data Catalog: Orbital elements of 156
  spectroscopic binaries (Mermilliod+, 2007)}, VizieR On-line Data Catalog:
  J/A+A/473/829. Originally published in: 2007A\&A...473..829M,
  \dodoi{10.26093/cds/vizier.34730829}

\bibitem[{{Mermilliod} {et~al.}(2007{\natexlab{b}}){Mermilliod}, {Andersen},
  {Latham}, \& {Mayor}}]{mermilliod07}
{Mermilliod}, J.-C., {Andersen}, J., {Latham}, D.~W., \& {Mayor}, M.
  2007{\natexlab{b}}, \aap, 473, 829, \dodoi{10.1051/0004-6361:20078007}

\bibitem[{{Moss} {et~al.}(2023){Moss}, {Kilic}, {Bergeron}, {Firgard}, \&
  {Brown}}]{Moss2023}
{Moss}, A., {Kilic}, M., {Bergeron}, P., {Firgard}, M., \& {Brown}, W. 2023,
  \mnras, 523, 5598, \dodoi{10.1093/mnras/stad1835}

\bibitem[{{North}(2014)}]{north14}
{North}, P. 2014, in Putting A Stars into Context: Evolution, Environment, and
  Related Stars, ed. G.~{Mathys}, E.~R. {Griffin}, O.~{Kochukhov}, R.~{Monier},
  \& G.~M. {Wahlgren}, 63--71.
\newblock \doarXiv{1309.7636}

\bibitem[{{O'Brien} {et~al.}(2024){O'Brien}, {Tremblay}, {Klein}, {Koester},
  {Melis}, {B{\'e}dard}, {Cukanovaite}, {Cunningham}, {Doyle}, {G{\"a}nsicke},
  {Gentile Fusillo}, {Hollands}, {McCleery}, {Pelisoli}, {Toonen},
  {Weinberger}, \& {Zuckerman}}]{O'Brien2024}
{O'Brien}, M.~W., {Tremblay}, P.~E., {Klein}, B.~L., {et~al.} 2024, \mnras,
  527, 8687, \dodoi{10.1093/mnras/stad3773}

\bibitem[{{Offner} {et~al.}(2023){Offner}, {Moe}, {Kratter}, {Sadavoy},
  {Jensen}, \& {Tobin}}]{Offner2023}
{Offner}, S.~S.~R., {Moe}, M., {Kratter}, K.~M., {et~al.} 2023, in Astronomical
  Society of the Pacific Conference Series, Vol. 534, Protostars and Planets
  VII, ed. S.~{Inutsuka}, Y.~{Aikawa}, T.~{Muto}, K.~{Tomida}, \& M.~{Tamura},
  275, \dodoi{10.48550/arXiv.2203.10066}

\bibitem[{Oliphant(2006)}]{Numpy_2006}
Oliphant, T. 2006, {NumPy}: A guide to {NumPy}, USA: Trelgol Publishing.
\newblock \url{http://www.numpy.org/}

\bibitem[{{{\"O}pik}(1924)}]{opik24}
{{\"O}pik}, E. 1924, Publications of the Tartu Astrofizica Observatory, 25, 1

\bibitem[{{Perren} {et~al.}(2023){Perren}, {Pera}, {Navone}, \&
  {V{\'a}zquez}}]{Perren2023}
{Perren}, G.~I., {Pera}, M.~S., {Navone}, H.~D., \& {V{\'a}zquez}, R.~A. 2023,
  \mnras, 526, 4107, \dodoi{10.1093/mnras/stad2826}

\bibitem[{{Raghavan} {et~al.}(2010){Raghavan}, {McAlister}, {Henry}, {Latham},
  {Marcy}, {Mason}, {Gies}, {White}, \& {ten Brummelaar}}]{raghavan10}
{Raghavan}, D., {McAlister}, H.~A., {Henry}, T.~J., {et~al.} 2010, \apjs, 190,
  1, \dodoi{10.1088/0067-0049/190/1/1}

\bibitem[{{Reipurth} {et~al.}(2014){Reipurth}, {Clarke}, {Boss}, {Goodwin},
  {Rodr{\'\i}guez}, {Stassun}, {Tokovinin}, \& {Zinnecker}}]{reipurth14}
{Reipurth}, B., {Clarke}, C.~J., {Boss}, A.~P., {et~al.} 2014, in Protostars
  and Planets VI, ed. H.~{Beuther}, R.~S. {Klessen}, C.~P. {Dullemond}, \&
  T.~{Henning}, 267, \dodoi{10.2458/azu\_uapress\_9780816531240-ch012}

\bibitem[{{Rekhi} {et~al.}(2024){Rekhi}, {Ben-Ami}, {Hallakoun}, {Shahaf},
  {Toonen}, \& {Rix}}]{Rekhi2024}
{Rekhi}, P., {Ben-Ami}, S., {Hallakoun}, N., {et~al.} 2024, arXiv e-prints,
  arXiv:2407.07048, \dodoi{10.48550/arXiv.2407.07048}

\bibitem[{{Roberts} {et~al.}(2024){Roberts}, {Pinsonneault}, {Johnson}, {Zinn},
  {Weinberg}, {Vrard}, {Tayar}, {Stello}, {Mosser}, {Johnson}, {Cao},
  {Stassun}, {Stringfellow}, {Serenelli}, {Mathur}, {Hekker}, {Garc{\'\i}a},
  {Elsworth}, \& {Corsaro}}]{Roberts2024}
{Roberts}, J.~D., {Pinsonneault}, M.~H., {Johnson}, J.~A., {et~al.} 2024,
  \mnras, 530, 149, \dodoi{10.1093/mnras/stae820}

\bibitem[{{Rosen} {et~al.}(2020){Rosen}, {Offner}, {Sadavoy}, {Bhandare},
  {V{\'a}zquez-Semadeni}, \& {Ginsburg}}]{rosen20}
{Rosen}, A.~L., {Offner}, S. S.~R., {Sadavoy}, S.~I., {et~al.} 2020, \ssr, 216,
  62, \dodoi{10.1007/s11214-020-00688-5}

\bibitem[{{Salpeter}(1955)}]{salpeter55}
{Salpeter}, E.~E. 1955, \apj, 121, 161, \dodoi{10.1086/145971}

\bibitem[{{Shahaf} {et~al.}(2023){Shahaf}, {Bashi}, {Mazeh}, {Faigler},
  {Arenou}, {El-Badry}, \& {Rix}}]{Shahaf2023}
{Shahaf}, S., {Bashi}, D., {Mazeh}, T., {et~al.} 2023, \mnras, 518, 2991,
  \dodoi{10.1093/mnras/stac3290}

\bibitem[{{Shahaf} {et~al.}(2024){Shahaf}, {Hallakoun}, {Mazeh}, {Ben-Ami},
  {Rekhi}, {El-Badry}, \& {Toonen}}]{Shahaf2024}
{Shahaf}, S., {Hallakoun}, N., {Mazeh}, T., {et~al.} 2024, \mnras, 529, 3729,
  \dodoi{10.1093/mnras/stae773}

\bibitem[{{Shahaf} \& {Mazeh}(2019)}]{Shahaf2019}
{Shahaf}, S., \& {Mazeh}, T. 2019, \mnras, 487, 3356,
  \dodoi{10.1093/mnras/stz1517}

\bibitem[{{Shahaf} {et~al.}(2017){Shahaf}, {Mazeh}, \& {Faigler}}]{shahaf17}
{Shahaf}, S., {Mazeh}, T., \& {Faigler}, S. 2017, \mnras, 472, 4497,
  \dodoi{10.1093/mnras/stx2257}

\bibitem[{{Shahaf} {et~al.}(2019){Shahaf}, {Mazeh}, {Faigler}, \&
  {Holl}}]{Shahaf2019b}
{Shahaf}, S., {Mazeh}, T., {Faigler}, S., \& {Holl}, B. 2019, \mnras, 487,
  5610, \dodoi{10.1093/mnras/stz1636}

\bibitem[{{Tang} {et~al.}(2014){Tang}, {Bressan}, {Rosenfield}, {Slemer},
  {Marigo}, {Girardi}, \& {Bianchi}}]{Tang_2014}
{Tang}, J., {Bressan}, A., {Rosenfield}, P., {et~al.} 2014, \mnras, 445, 4287,
  \dodoi{10.1093/mnras/stu2029}

\bibitem[{{Tremblay} {et~al.}(2024){Tremblay}, {B{\'e}dard}, {O'Brien},
  {Munday}, {Elms}, {Gentillo Fusillo}, \& {Sahu}}]{Tremblay2024}
{Tremblay}, P.-E., {B{\'e}dard}, A., {O'Brien}, M.~W., {et~al.} 2024, \nar, 99,
  101705, \dodoi{10.1016/j.newar.2024.101705}

\bibitem[{{Tremblay} {et~al.}(2016){Tremblay}, {Cummings}, {Kalirai},
  {G{\"a}nsicke}, {Gentile-Fusillo}, \& {Raddi}}]{tremblay16}
{Tremblay}, P.-E., {Cummings}, J., {Kalirai}, J.~S., {et~al.} 2016, \mnras,
  461, 2100, \dodoi{10.1093/mnras/stw1447}

\bibitem[{{van Biesbroeck}(1916)}]{biesbroeck1916}
{van Biesbroeck}, G. 1916, \aj, 29, 173, \dodoi{10.1086/104155}

\bibitem[{{Van der Swaelmen} {et~al.}(2017){Van der Swaelmen}, {Boffin},
  {Jorissen}, \& {Van Eck}}]{swaelmen17}
{Van der Swaelmen}, M., {Boffin}, H.~M.~J., {Jorissen}, A., \& {Van Eck}, S.
  2017, \aap, 597, A68, \dodoi{10.1051/0004-6361/201628867}

\bibitem[{{{van der} Walt} {et~al.}(2011){{van der} Walt}, {Colbert}, \&
  {Varoquaux}}]{Numpy_2011}
{{van der} Walt}, S., {Colbert}, S.~C., \& {Varoquaux}, G. 2011, Computing in
  Science Engineering, 13, 22, \dodoi{10.1109/MCSE.2011.37}

\bibitem[{{Virtanen} {et~al.}(2020){Virtanen}, {Gommers}, {Oliphant},
  {Haberland}, {Reddy}, {Cournapeau}, {Burovski}, {Peterson}, {Weckesser},
  {Bright}, {van der Walt}, {Brett}, {Wilson}, {Millman}, {Mayorov}, {Nelson},
  {Jones}, {Kern}, {Larson}, {Carey}, {Polat}, {Feng}, {Moore}, {Vand erPlas},
  {Laxalde}, {Perktold}, {Cimrman}, {Henriksen}, {Quintero}, {Harris},
  {Archibald}, {Ribeiro}, {Pedregosa}, {van Mulbregt}, \& {SciPy 1. 0
  Contributors}}]{Virtanen_2020}
{Virtanen}, P., {Gommers}, R., {Oliphant}, T.~E., {et~al.} 2020, Nature
  Methods, 17, 261, \dodoi{10.1038/s41592-019-0686-2}

\bibitem[{{Wagg} {et~al.}(2024){Wagg}, {Breivik}, {Renzo}, \&
  {Price-Whelan}}]{Wagg2024}
{Wagg}, T., {Breivik}, K., {Renzo}, M., \& {Price-Whelan}, A.~M. 2024, arXiv
  e-prints, arXiv:2409.04543, \dodoi{10.48550/arXiv.2409.04543}

\bibitem[{Wagg {et~al.}(2025)Wagg, Breivik, Renzo, \& Price-Whelan}]{Wagg2025}
Wagg, T., Breivik, K., Renzo, M., \& Price-Whelan, A.~M. 2025, Journal of Open
  Source Software, 10, 7400, \dodoi{10.21105/joss.07400}

\bibitem[{{Yamaguchi} {et~al.}(2024{\natexlab{a}}){Yamaguchi}, {El-Badry},
  {Rees}, {Shahaf}, {Mazeh}, \& {Andrae}}]{Yamaguchi2024b}
{Yamaguchi}, N., {El-Badry}, K., {Rees}, N.~R., {et~al.} 2024{\natexlab{a}},
  \pasp, 136, 084202, \dodoi{10.1088/1538-3873/ad6809}

\bibitem[{{Yamaguchi} {et~al.}(2024{\natexlab{b}}){Yamaguchi}, {El-Badry},
  {Fuller}, {Latham}, {Cargile}, {Mazeh}, {Shahaf}, {Bieryla}, {Buchhave}, \&
  {Hobson}}]{Yamaguchi2024}
{Yamaguchi}, N., {El-Badry}, K., {Fuller}, J., {et~al.} 2024{\natexlab{b}},
  \mnras, 527, 11719, \dodoi{10.1093/mnras/stad4005}

\end{thebibliography}
\bibliographystyle{aasjournal}



\end{document}